# Impact of resource availability and conformity effect on sustainability of common-pool resources

**Authors:** Chengyi Tu[1,2*], Renfei Chen[3], Ying Fan[4], Xuwei Pan[2]

[1]Department of Environmental Science, Policy, and Management, University of California, Berkeley; Berkeley, 94720, USA.

[2]School of Economics and Management, Zhejiang Sci-Tech University; Hangzhou, 310018, China.

[3]School of Life Science, Shanxi Normal University; Taiyuan, 030000, China.

[4]College of Geography and Environment, Shandong Normal University; Jinan, 250358, China.

*Corresponding author. Email: chengyitu1986@gmail.com

# Abstract

Sustainability of common-pool resources hinges on the interplay between human and environmental systems. However, there is still a lack of a novel and comprehensive framework for modelling extraction of common-pool resources and cooperation of human agents that can account for different factors that shape the system behavior and outcomes. In particular, we still lack a critical value for ensuring resource sustainability under different scenarios. In this paper, we present a novel framework for studying resource extraction and cooperation in human-environmental systems for common-pool resources. We explore how different factors, such as resource availability and conformity effect, influence the players' decisions and the resource outcomes. We identify critical values for ensuring resource sustainability under various scenarios. We demonstrate the observed phenomena are robust to the complexity and assumptions of the models and discuss implications of our study for policy and practice, as well as the limitations and directions for future research.

# Introduction

The management of common-pool resources (CPR) such as fisheries, forests, or water, presents a collective action social dilemma[1-3]. Individual rationality can lead to collective irrationality and resource degradation, a phenomenon known as the tragedy of the commons[4,5]. To prevent this, users must cooperate and coordinate their extraction and management of the resource to ensure its long-term sustainability. However, achieving and maintaining cooperation is often challenging due to various factors such as the number and diversity of users, the characteristics and dynamics of the resource, environmental uncertainty and variability, and the social norms and institutions that regulate resource use[4,6]. Therefore, a comprehensive framework that can capture the complex interactions and feedbacks between human and environmental systems (HES) is essential for understanding and managing CPR.

HES for CPR is comprised of two interdependent components[7-9]: human agents and natural resources. Human agents are individuals or groups that consume, manage, or affect the resource. They have different preferences, beliefs, and strategies, and can influence each other through social networks or institutions. Natural resources are biophysical entities that provide benefits or services to human agents. They have different properties such as renewability, scarcity, or spatial distribution, and can change over time according to their natural dynamics or human interventions. The interactions between human agents and natural resources can produce various outcomes such as resource sustainability or depletion, cooperation emergence or collapse, or social welfare improvement or decline. These outcomes can in turn alter the behavior and preferences of human agents and the state and dynamics of natural resources, creating feedback loops that drive the evolution of HES.

However, there is a gap in the literature on a novel and comprehensive HES framework for modelling extraction of CPR and cooperation of human agents that can account for different factors that shape system behavior and outcomes[10-14]. In particular, players may make decisions that depend on resource availability and conformity effect in real-world scenarios. The resource availability reflects environmental conditions and collective actions of players, influencing their strategies and payoffs[15,16]. The conformity effect reflects social influence and peer pressure experienced by players, influencing their strategies and payoffs[17,18]. Moreover, a key challenge for CPR management is identifying a critical value that guarantees resource sustainability under different scenarios and parameter configurations[19-21]. This critical value represents the threshold of the initial condition ensuring system convergence to a sustainable state regardless of player strategies and payoffs. This critical value has significant implications for policy and practice as it can inform design and implementation of effective interventions to prevent or reverse resource degradation[9,22-24].

In this paper, we present a novel framework that models the coupled dynamics of resource extraction and cooperation in HES. We use agent-based modelling as a methodological tool to represent heterogeneous human agents and their interactions with each other and with the resource pool. We use game theory as a conceptual tool to analyze the strategic selection and payoffs of human agents in different scenarios. We explore how different factors, such as resource availability, conformity effect, and their balance, influence the players' decisions and the resource outcomes. We identify critical values for ensuring resource sustainability under various scenarios and parameter configurations. We demonstrate that the observed phenomena are robust to the complexity and assumptions of the models and discuss the implications of our study for policy and practice, as well as the limitations and directions for future research.

# Results

## Minimal model

We consider a model of resource extraction with a common pool that has a resource volume $R$ that follows a logistic function $TR\left(1-\frac{R}{K}\right)$ where $T>0$ is the natural growth rate, $K$ is the carrying capacity[25-27] and is exploited by $N$ players[28,29]. Each player can adopt either a cooperation, which extracts a sustainable amount of resource, or a defection, which extracts a larger and unsustainable amount. The payoffs for cooperators and defectors are $U_C = Re_C$ and $U_D = Re_D$, respectively, where $e_C$ and $e_D$ are the efforts exerted by each strategy. We model the dynamic evolution of the fraction of cooperators as replicator rule with a greed parameter $w$, where each player compares its payoff with payoff of one of its neighbor and then selects the cooperation or defection[30]. The greed parameter reflects how much a player prefers to extract more resource by defection than the sustainable level by cooperation. When $w>0$, the fraction of cooperators $x$ decreases and the fraction of defectors $1-x$ increases. When $w<0$, the opposite occurs. When $w=0$, the fraction of cooperators remains constant over time. Moreover, the larger the absolute value of $w$, the faster the evolution of the players' strategy. By coupling the evolutionary dynamics of resource and players' strategies described above (see Fig. 1 a-c), the macroscopic equations that approximate the microscopic update of the players harvesting the resource are (see Methods):

$$\begin{cases} \dfrac{dR(t)}{dt} &= T\left(R(t)\left(1-\dfrac{R(t)}{K}\right)-R(t)\left(x(t)\hat{e}_C+(1-x(t))\hat{e}_D\right)\right) \\ \dfrac{dx(t)}{dt} &= -w(t)R(t)x(t)(1-x(t)) \end{cases} \quad (1)$$

where $0 \leq R(t) \leq 1$ and $0 \leq x(t) \leq 1$ are the resource volume and the fraction of cooperators at time $t$, $\hat{e}_C = \dfrac{Ne_C}{T}, \hat{e}_D = \dfrac{Ne_D}{T}$ are normalized extraction rates of cooperation and defection where $0 < \hat{e}_C < 1 < \hat{e}_D$ and $K$ can set to 1 by normalizing $R$ for simplicity. The microscopic update reflects the individual-level strategy switching of the players and the subsequent resource update, while the macroscopic ODE captures the population-level dynamics of the fraction of cooperators and resource. We compare its microscopic update and macroscopic ODE with different parameter configurations $\hat{e}_C = 0.3, \hat{e}_D = 1.1$, $\hat{e}_C = 0.3, \hat{e}_D = 1.5$, $\hat{e}_C = 0.7, \hat{e}_D = 1.1$ and $\hat{e}_C = 0.7, \hat{e}_D = 1.5$ for $w = -1, 0, 1$ and find consistent results between them (see Supplementary Fig. 1-3).

We derive the steady-state solutions of the coupled system and assessed their stability analytically. The CPR sustainability of the system depends on the parameter configuration and initial condition (see Methods): Scenario $w<0$:

the system is bi-stable and $x_c^M = \frac{-1+\hat{e}_D}{-\hat{e}_C+\hat{e}_D}$ is a critical value for ensuring the CPR sustainability, i.e., if $x_0 > x_c^M$, then the system is assured to sustain the resource, regardless of the initial condition of $R_0$; Scenario $w=0$: if $x_0 < x_c^M$, the system has only one unstable solution that depletes the resource; if $x_0 > x_c^M$, the system has only one stable solution that sustains the resource; Scenario $w>0$: the system always depletes the resource irrespective of the parameter configuration and initial condition. In summary, if the greed parameter $w$ is non-positive and initial condition $x_0 > x_c^M$, the system sustains the resource; otherwise, the system depletes the resource (see Fig. 1 d-f and Supplementary Fig. 4-6).

The minimal model is a simplified model that assumes a constant greed parameter, which is difficult to measure in real-world scenarios. Furthermore, the players may make decisions based on other factors besides the standard replicator dynamics, such as resource availability and conformity effect. Therefore, we explore a variety of more complex models that alter the minimal model in different ways. We show that these models produce qualitatively consistent results with the minimal model, indicating that the observed phenomena are not dependent on the simplicity or assumptions of the minimal model.

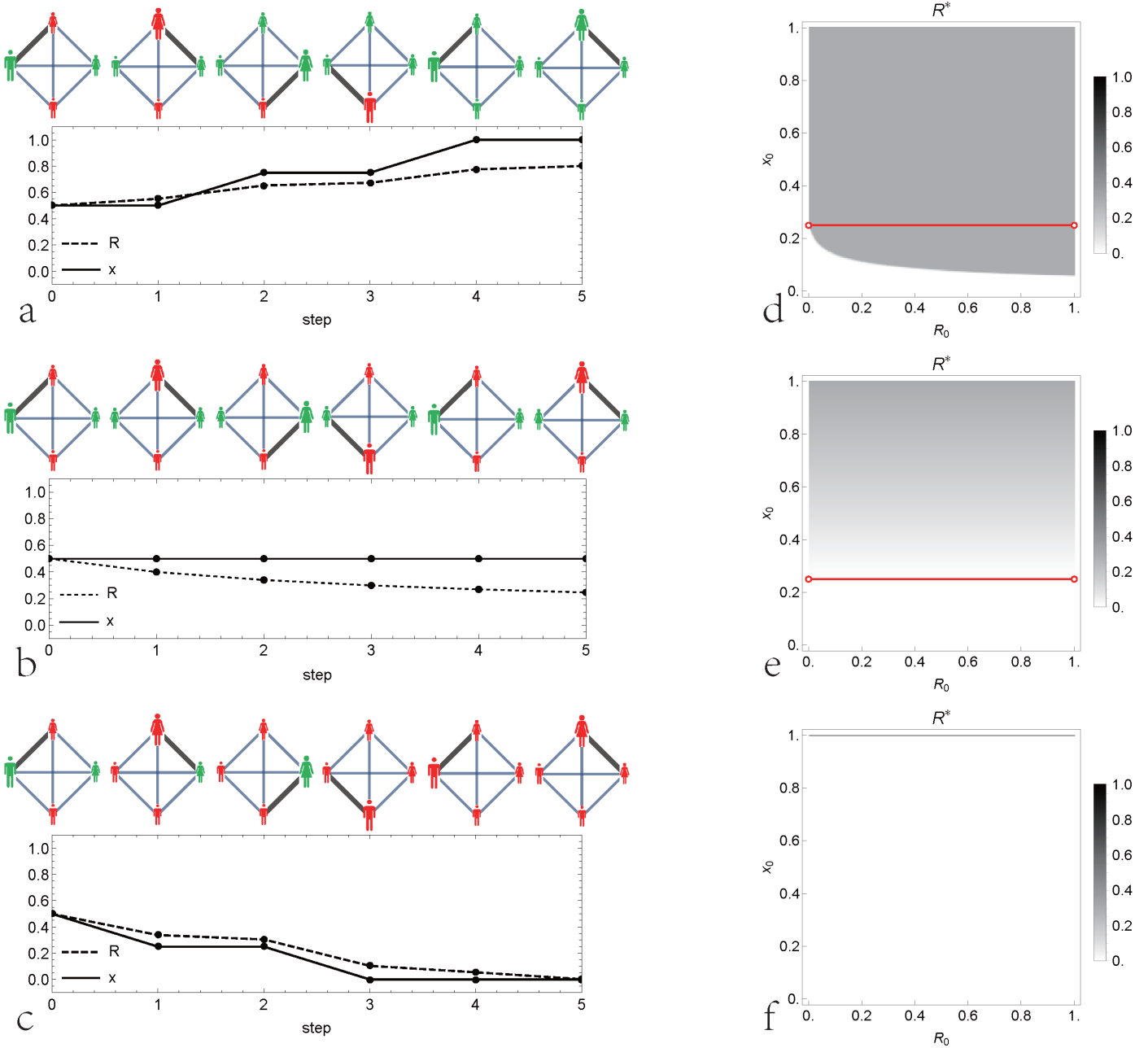

**Fig. 1 | A schematic of the minimal model.** (**a-c**) Evolution of the minimal model with the negative, zero and positive greed parameter $w$. At each step the active player (the highlighted node) compares its payoff with one of its neighbors' where cooperators/defectors are shown in green/red. (**d-f**) Density plot of stable equilibrium $R^*$ for the minimal model with the negative, zero and positive greed parameter $w$. The other parameters are growth rate $T = 2$, normalized extraction parameter $\hat{e}_C = 0.7, \hat{e}_D = 1.1$.

## Resource-driven model

In real-world scenarios, the players may make decisions that depend on the resource availability in addition to the replicator dynamics[15,16]. The resource availability reflects the environmental conditions and the collective actions of the players, and may influence the players' strategies and payoffs. Therefore, we introduce a resource-driven greed parameter that captures the effect of resource availability on the players' decisions. This parameter determines how much a player is willing to cooperate or defect based on the current level of resource in the system, i.e., a player is more likely to cooperate when the resource is scarce and defect when the resource is abundant for CPR sustainability. We explore how this parameter affects the dynamics and outcomes of the system, and show that it can lead to different regimes of stable solutions and CPR sustainability depending on the parameter configuration and initial condition of the system.

We employ a linear function of the resource as the greed parameter, i.e., $w(t) = 2R(t) - 1$ to capture this behavior (see Method). We compare its microscopic update and macroscopic ODE with different parameter configurations $\hat{e}_C = 0.3, \hat{e}_D = 1.1$, $\hat{e}_C = 0.3, \hat{e}_D = 1.5$, $\hat{e}_C = 0.7, \hat{e}_D = 1.1$ and $\hat{e}_C = 0.7, \hat{e}_D = 1.5$ and find consistent results between them (see Supplementary Fig. 7). We obtain the analytical solutions for the steady-state behaviour of the coupled system and evaluate their stability. The CPR sustainability of the system depends on the parameter configuration and initial condition (see Methods): $x_c^R = x_c^M = \dfrac{-1 + \hat{e}_D}{-\hat{e}_C + \hat{e}_D}$ is a critical value for ensuring CPR sustainability for any initial condition $\{R_0, x_0\}$, i.e., if initial condition $x_0 > x_c^R$, then the system is assured to sustain the resource, regardless of the initial condition of $R_0$; otherwise, it may deplete the resource (see Fig. 2 a-d and Supplementary Fig. 8).

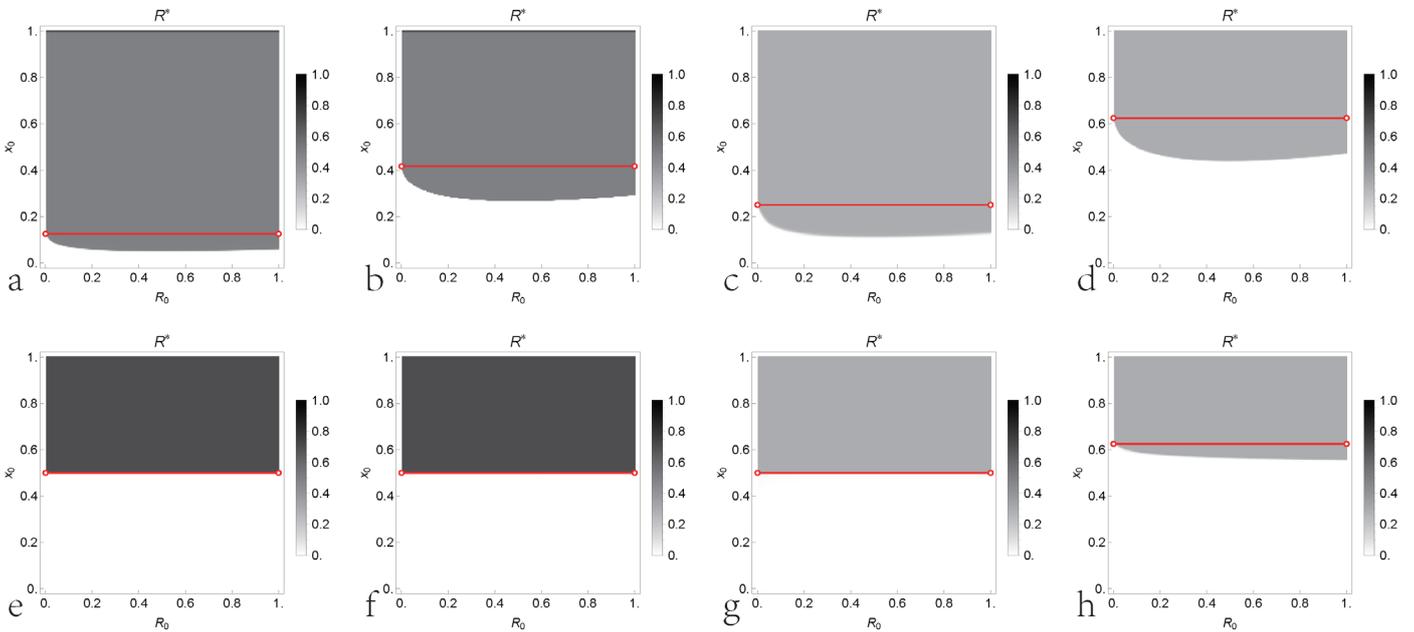

**Fig. 2 | The effect of resource availability and conformity effect on the stable equilibrium $R^*$ of the coupled system.** (a-d) Density plot of stable equilibrium $R^*$ for the resource-driven model with the parameter configurations $\hat{e}_C = 0.3, \hat{e}_D = 1.1$, $\hat{e}_C = 0.3, \hat{e}_D = 1.5$, $\hat{e}_C = 0.7, \hat{e}_D = 1.1$ and $\hat{e}_C = 0.7, \hat{e}_D = 1.5$. (e-h) Density plot of stable equilibrium $R^*$ for the conformity-driven model with the same parameter configurations. In each panel, the critical value

for any initial condition of $R_0$ is denoted by the red line.

## Conformity-driven model

In real-world scenarios, the players may make decisions that depend on the conformity effect in addition to the replicator dynamics[17,18]. The conformity effect reflects the social influence and the peer pressure that the players experience, and may influence the players' strategies and payoffs. Therefore, we introduce a conformity-driven greed parameter that captures the effect of conformity on the players' decisions. This parameter determines how much a player is willing to cooperate or defect based on the fraction of cooperators in the system, i.e., a player is more likely to cooperate when most players are cooperators and defect when most players are defectors. We explore how this parameter affects the dynamics and outcomes of the system, and show that it can lead to different regimes of stable solutions and CPR sustainability depending on the parameter configuration and initial condition of the system.

We employ a linear function of the fraction of cooperators as greed parameter, i.e., $w(t) = 1 - 2x(t)$ to capture this behavior (see Method). We compare its microscopic update and macroscopic ODE with different parameter configurations $\hat{e}_C = 0.3, \hat{e}_D = 1.1$, $\hat{e}_C = 0.3, \hat{e}_D = 1.5$, $\hat{e}_C = 0.7, \hat{e}_D = 1.1$ and $\hat{e}_C = 0.7, \hat{e}_D = 1.5$ and find consistent results between them (see Supplementary Fig. 9). We obtain the analytical solutions for the steady-state behaviour of the coupled system and evaluate their stability. The CPR sustainability of the system depends on the parameter configuration and initial condition (see Methods): $x_c^C = \max\{x_c^M, 1/2\}$ is a critical value for ensuring CPR sustainability for any initial condition $\{R_0, x_0\}$, i.e., if initial condition $x_0 > x_c^C$, then the system is assured to sustain the resource, regardless of the initial condition of $R_0$; otherwise, it may deplete the resource (see Fig. 2 e-h and Supplementary Fig. 10).

## Resource and conformity-driven model

In more real-world scenarios, the players may make decisions that depend on the both resource availability and conformity effect in addition to the replicator dynamics. Therefore, we introduce a resource and conformity-driven greed parameter that captures both the effect of resource and conformity on the players' decisions. This parameter determines how much a player is willing to cooperate or defect based on the current level of resource and fraction of cooperators in the system. We explore how this parameter affects the dynamics and outcomes of the system, and show that it can lead to different regimes of stable solutions and CPR sustainability depending on the parameter configuration and initial condition of the system.

We employ a linear function of the resource and fraction of cooperators as greed parameter, i.e., $w = (1-c)R(t) + (-1-c)x(t) + c$ to capture this behavior (see Method). We compare its microscopic update and macroscopic ODE with different parameter configurations $\hat{e}_C = 0.3, \hat{e}_D = 1.1$, $\hat{e}_C = 0.3, \hat{e}_D = 1.5$, $\hat{e}_C = 0.7, \hat{e}_D = 1.1$ and $\hat{e}_C = 0.7, \hat{e}_D = 1.5$ for $c = -0.75, -0.25, 0.25, 0.75$ and find consistent results between them (see Supplementary Fig. 11-14). We obtain the analytical solutions for the steady-state behaviour of the coupled system and evaluate their stability. The CPR sustainability of the system depends on the parameter configuration and initial condition (see Methods):

$$x_c^{RC} = \frac{1-c}{2}x_c^R + \frac{-1-c}{-2} = x_c^C = \frac{1-c}{2}x_c^M + \frac{-1-c}{-2}\max\{x_c^M, 1/2\}$$

is a critical value for ensuring CPR sustainability for any initial condition $\{R_0, x_0\}$, i.e., if initial condition $x_0 > x_c^{RC}$, then the system is assured to sustain the resource, regardless of the initial condition of $R_0$; otherwise, it may deplete the resource (see Fig. 3 and Supplementary Fig. 15-18).

When the parameter $c = -1$, indicating that players' decisions are solely influenced by resource availability, the resource and conformity-driven model simplifies to the resource-driven model (see Fig. 2 c). Conversely, when the parameter $c = 1$, indicating that players' decisions are solely influenced by the conformity effect, the resource and conformity-driven model simplifies to the conformity-driven model (see Fig. 2 g). These observations highlight the adaptability of the resource and conformity-driven model to different decision-making influences.

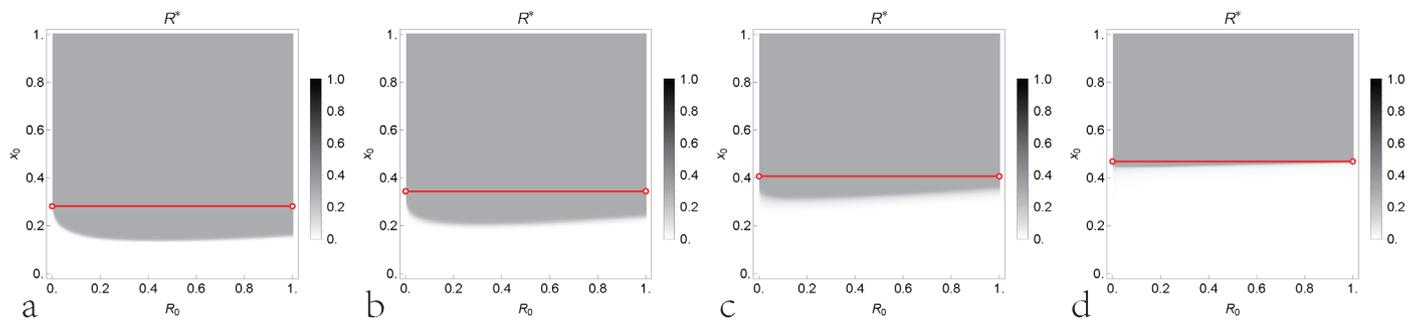

**Fig. 3 | The effect of both resource availability and conformity effect on the stable equilibrium $R^*$ of the coupled system.** (a-d) Density plot of stable equilibrium $R^*$ for the resource and conformity-driven model with the parameter configuration $\hat{e}_C = 0.7, \hat{e}_D = 1.1$ for $c = -0.75, -0.25, 0.25, 0.75$ respectively. In each panel, the critical value for any initial condition of $R_0$ is denoted by the red line.

## Comparison of critical value for different models

Finally, we investigate the effect of the parameter configuration $\hat{e}_C, \hat{e}_D$ on the critical value for different models

$x_c^M, x_c^R, x_c^C, x_c^{RC}$. The critical value is defined as a threshold value that ensures the CPR sustainability for any initial condition $\{R_0, x_0\}$, meaning that the if initial condition exceeds this value, then the system is assured to sustain the resource; otherwise, it may deplete the resource. For the minimal model: If the greed parameter $w \leq 0$ (see Fig. 4 a), the critical value is $x_c^M = \dfrac{-1+\hat{e}_D}{-\hat{e}_C+\hat{e}_D}$ and increases monotonically with the parameter configuration $\hat{e}_C, \hat{e}_D$. As they approach their lower bound $\hat{e}_C \to 0, \hat{e}_D \to 1$, $x_c^M$ approaches its lower bound 0; as $\hat{e}_C \to 1$ approaches its upper bound for any feasible $\hat{e}_D$, $x_c^M$ approaches its upper bound 1; if the greed parameter $w > 0$, the system depletes the resource. Consequently, there is no critical value for CPR sustainability. For the resource-driven model, the critical value is the same as minimal model with greed parameter $w \leq 0$, i.e., $x_c^R = x_c^M$. For the conformity-driven model (see Fig. 4 b): If the given parameter configuration $\hat{e}_C, \hat{e}_D$ lie in the right-top region of the linear function $\hat{e}_D = -\hat{e}_C + 2$, then the critical value is the same as the minimal model $x_c^C = x_c^M$; If they lie in the left-bottom region of the linear function, then the critical value is $x_c^C = 0.5$. For the resource and conformity-driven model (see Fig. 4 c-f): If the given parameter configuration $\hat{e}_C, \hat{e}_D$ lie in the right-top region of the linear function, then the critical value is the same as minimal model $x_c^{RC} = x_c^M$; If they lie in the left-bottom region of the linear function, then the critical value is $x_c^{RC} = \dfrac{1-c}{2} x_c^M + \dfrac{-1-c}{-2} 0.5$.

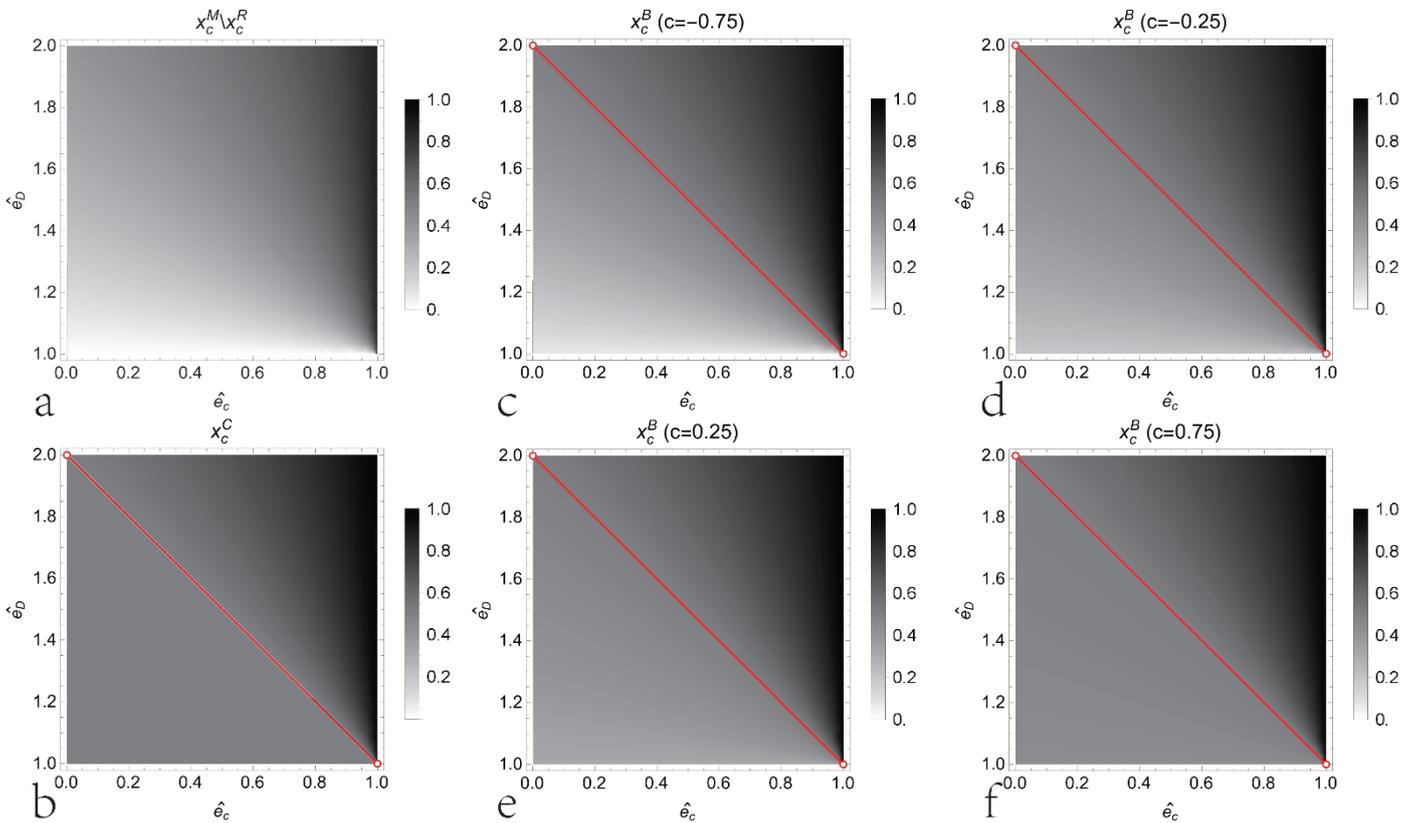

**Fig. 4 | The effect of different models on the critical value of the system.** (a) Density plot of critical value $x_c^M$ or $x_c^R$ for

the minimal model or the resource-driven model. (**b**) Density plot of critical value $x_c^C$ for the conformity-driven model. (**c-f**) Density plot of critical value $x_c^B$ for the resource and conformity-driven model with parameter $c = -0.75, -0.25, 0.25, 0.75$ respectively. In each panel, the linear function $\hat{e}_D = -\hat{e}_C + 2$ is denoted by the red line.

To summarize, the critical value of all models is equal, i.e., $x_c^M = x_c^R = x_c^C = x_c^{RC}$, when the parameter configuration $\hat{e}_C, \hat{e}_D$ falls in the right-top region of the linear function $\hat{e}_D = -\hat{e}_C + 2$ (see Tal. 1). This indicates that the CPR sustainability is not affected by any effects driven by the greed parameter. Conversely, the critical value of the conformity-driven model is higher than that of the resource and conformity-driven model, which is higher than that of the resource-driven model which is equal to that of the minimal model, i.e., $x_c^C > x_c^{RC} > x_c^R = x_c^M$, when the parameter configuration $\hat{e}_C, \hat{e}_D$ falls in the left-bottom region of the linear function (see Tal. 1). This indicates that the CPR sustainability is negatively affected by the conformity effects driven by the greed parameter, whereas the resource availability has no effect.

Table 1. Critical values for CPR sustainability under different models.

| Model with parameters | | Critical value |
|---|---|---|
| minimal model | $w > 0$ | inexistence |
| | $w \leq 0$ | $x_c^M = \dfrac{-1 + \hat{e}_D}{-\hat{e}_C + \hat{e}_D}$ |
| resource-driven model | | $x_c^R = x_c^M$ |
| conformity-driven model | $\hat{e}_C, \hat{e}_D$ lie in the right-top region of the linear function $\hat{e}_D = -\hat{e}_C + 2$ | $x_c^C = x_c^M$ |
| | $\hat{e}_C, \hat{e}_D$ lie in the left-bottom region of the linear function $\hat{e}_D = -\hat{e}_C + 2$ | $x_c^C = 0.5$ |
| resource and conformity-driven model | $\hat{e}_C, \hat{e}_D$ lie in the right-top region of the linear function $\hat{e}_D = -\hat{e}_C + 2$ | $x_c^{RC} = x_c^M$ |
| | $\hat{e}_C, \hat{e}_D$ lie in the left-bottom region of the linear function $\hat{e}_D = -\hat{e}_C + 2$ | $x_c^{RC} = \dfrac{1-c}{2} x_c^M + \dfrac{-1-c}{-2} 0.5$ |

These results can be explained as follows: Both resource availability and conformity effect are important for CPR sustainability, as they represent real-world situations that can override the replicator rule that favours defection for higher individual payoff. For the resource-driven model, the players make decisions that depend on the resource availability. A player is more likely to cooperate when the resource is scarce and defect when the resource is abundant for CPR sustainability, for the sake of CPR sustainability. This induces all players to consider CPR sustainability more and eventually override the replicator rule. For the conformity-driven model, the players make decisions that depend on the conformity effect, which can induce players to follow the majority behavior, regardless of its consequences for the resource or their own payoffs. If both parameter configuration $\hat{e}_C, \hat{e}_D$ are small (the linear function $\hat{e}_D = -\hat{e}_C + 2$) so that the CPR extraction is weak: When the initial condition $x_0 < 0.5$, i.e., the fraction of defectors exceeds the fraction of cooperators, the conformity effect will increase the fraction of defectors and eventually deplete the CPR. When the initial condition $x_0 > 0.5$, i.e., the fraction of cooperators exceeds the fraction of defectors, the conformity effect will increase the fraction of cooperators, and possibly counteract the effect of the replicator rule. However, if both parameter configurations are large, so that the CPR extraction is strong, the conformity-driven model becomes similar to the minimal model. For the resource and conformity-driven model, it is a balance of the resource-driven model and the conformity-driven model and the critical linear function of the conformity-driven model still exists.

# Discussion

This paper presents a novel framework for studying resource extraction and cooperation in human-environmental systems (HES) for common-pool resources (CPR). The paper uses microscopic update based on agent-based modelling and macroscopic ODE based on game theory to explore how different factors, such as resource availability, conformity effect, and their balance, influence the players' decisions and the resource outcomes. The paper identifies critical values for ensuring the CPR sustainability under different scenarios and parameter configurations and find that the critical value of all models is equal, i.e., $x_c^M = x_c^R = x_c^C = x_c^{RC}$, when the parameter configuration $\hat{e}_C, \hat{e}_D$ falls in the right-top region of the linear function $\hat{e}_D = -\hat{e}_C + 2$ and the critical value of the conformity-driven model is higher than that of the resource and conformity-driven model, which is higher than that of the resource-driven model which is equal to that of the minimal model, i.e., $x_c^C > x_c^{RC} > x_c^R = x_c^M$, when the parameter configuration $\hat{e}_C, \hat{e}_D$ falls in the left-bottom region of the linear function. The authors demonstrate that the observed phenomena are robust to the complexity and assumptions of the models.

The paper considers a linear function for the greed parameter that depends on resource availability, conformity effect and their balance. It would be worthwhile to explore other functional forms or nonlinearities that may capture more realistic behaviors of the players. To extend our analysis to more realistic cases, we also consider a quadratic form, i.e., $w = aR^2 + bR + c$ for the resource-driven model, $w = ax^2 + bx + c$ for the conformity-driven model and $w = aR^2 + bR + cx^2 + dx + e$ for the resource and conformity-driven model (see Supplementary Notes). The analytical investigation of the properties and stability of the solution is challenging. Therefore, we use numerical simulations to examine the dynamic behavior of the system. Our results reveal that the system displays different patterns depending on the value of the parameter configuration $\hat{e}_C, \hat{e}_D$ and initial condition $\{R_0, x_0\}$ and parameter of quadratic form, but they are qualitatively similar to those observed in the linear case (see Supplementary Fig. 19-21).

To account for the effects of social interactions on the players' decisions, we also consider two alternative scenarios where the players are not fully connected but only linked to a subset of other players through different types of social networks (see Supplementary Notes). The first scenario is a Barabasi-Albert (BA) network, which is characterized by a highly heterogeneous degree distribution, meaning that the number of connections at each player in the network varies widely. This structure is often observed in real-world social networks, as it can emerge from a preferential attachment mechanism, where new nodes tend to connect to existing nodes with higher degrees. In this scenario, a few players have an extremely large number of neighbours, while most players have very few neighbours. The second scenario is a small-world (SW) network, which is characterized by a high clustering coefficient and a low average path length, meaning that each player can be reached from every other player through a small number of steps, and that players that share a common neighbour are likely to be connected as well. This structure captures the features of many natural and social systems, where local interactions and global connectivity coexist. We compare the outcomes of the microscopic update with different social networks with macroscopic ODE and find consistent results between them, so our main conclusions are robust and do not depend significantly on the type of social network connecting the players (see Supplementary Fig. 22-24).

The paper makes several important contributions to the literature on CPR management and HES modelling. First, the paper provides a comprehensive framework that can capture the complex interactions and feedbacks between human agents and natural resources in CPR settings. The framework can account for different factors that shape the system behavior and outcomes. Second, the paper introduces a novel concept of a greed parameter that reflects how players adjust their strategies based on resource availability or conformity effect, or both. These models capture more realistic behaviors of the players and allow for more flexibility and diversity in the system. Third, the paper analyzes the steady-state solutions and stability of the coupled system analytically and numerically, and provides insights into the conditions for the CPR sustainability.

The paper has implications for policy and practice, as it provides insights into how to design effective interventions and incentives to promote the CPR sustainability in HES. The paper suggests that managing both resource availability and conformity effect is crucial for achieving desirable outcomes, and that different scenarios may require different strategies depending on the parameter configurations and initial conditions of the system. For example, the paper shows that resource availability has no effect on CPR sustainability, whereas conformity effect has a negative effect. This suggests that policies that enhance resource availability may not be sufficient to prevent overexploitation and degradation of CPR, whereas policies that promote cooperation and coordination among users may be more effective. This framework and analysis can be applied to CPR and social dilemmas, such as climate change, public goods, or collective action.

The paper has some limitations that could be addressed in future work. First, the paper assumes a homogeneous population of players who have the same extraction rates and payoffs for each strategy. It would be interesting to explore how heterogeneity among players affects the dynamics and outcomes of the system. For example, how does inequality in extraction rates or payoffs influence the players' decisions and the resource dynamics? How does diversity in preferences or beliefs affect the cooperation level and the CPR sustainability? Second, the paper focuses on a static setting where the parameters are fixed over time. It would be relevant to investigate how temporal variations or feedbacks in the parameters influence the evolution of the system. For example, how does stochasticity or uncertainty in resource growth rate or extraction rates affect the system stability and resilience? How does adaptation or learning in greed parameter or extraction rates affect the system dynamics and outcomes?

In conclusion, the paper provides a novel and comprehensive framework for modelling resource extraction and cooperation in HES for CPR, demonstrates that different factors play important roles in shaping the system behaviour and outcomes and identifies critical values for ensuring CPR sustainability under different scenarios and parameter configurations. The paper demonstrates that the observed phenomena are robust to the complexity and assumptions of the models and discuss the implications of their study for policy and practice, as well as the limitations and directions for future research.

# Methods

## Evolutionary dynamics of resource

We develop a model of a dynamic resource pool that is exploited by players who can choose to cooperate or defect. Cooperation entails a sustainable extraction of resource, while defection entails a larger and unsustainable extraction. The resource pool has a volume that follows a logistic function $TR(t)\left(1-\frac{R(t)}{K}\right)$ with a natural growth rate $T>0$ and a carrying capacity $K$. The extraction rates for cooperators and defectors are $e_C, e_D$, respectively, with $0 < Ne_C < T < Ne_D$. The fraction of cooperators in the system is $x = \frac{N_C}{N}$, where $N_C$ is the number of cooperators (and the fraction of defectors is $1-x$). The total extraction rate by all players is

$E = N_C e_C + (N - N_C) e_D = N(x e_C + (1-x) e_D)$. The differential equation for the resource volume is then:

$$
\begin{aligned}
\frac{dR(t)}{dt} &= TR(t)\left(1 - \frac{R(t)}{K}\right) - R(t)E = TR(t)\left(1 - \frac{R(t)}{K}\right) - NR(t)\left(x(t)e_C + (1-x(t))e_D\right) \\
&= T\left(R(t)\left(1 - \frac{R(t)}{K}\right) - R(t)\left(x(t)\hat{e}_C + (1-x(t))\hat{e}_D\right)\right)
\end{aligned}
\qquad (2)
$$

where $\hat{e}_C = \frac{Ne_C}{T}, \hat{e}_D = \frac{Ne_D}{T}$ are normalized extraction parameters and $K = 1$ by normalizing the resource volume $R$ between 0 and 1 for simplicity.

## Evolutionary dynamics of players' strategies

We derive a mean-field equation for the evolution of the players' strategy (see Supplementary Notes), which is:

$$\frac{dx(t)}{dt} = T^{D \to C} - T^{C \to D} \qquad (3)$$

where $T^{D \to C}, T^{C \to D}$ is transition probability for Master Equation. We start from the simplest case of the replicator rule, which models the players' strategy as follows: 1) a player $i$ is randomly chosen; 2) a neighbor of $i$ is randomly chosen, $j$; 3) the strategy of player $i$ to switch to the strategy of player $j$ with probability $p = \frac{1}{2} + \frac{w}{2}\frac{U_j - U_i}{\Delta U_{max}}$ where $U_i$ and $U_j$ are the payoffs of players $i$ and $j$, respectively, $\Delta U_{max}$ is the maximum possible payoff difference that ensures $0 \le p \le 1$ ($\Delta U_{max} = \max|U_D - U_C| = e_D - e_C$ when $R = 1$) and $-1 \le w \le 1$ is the greed parameter[28]. In this setting, the probabilities to switch strategy from defection to cooperation, and vice versa are

$p^{D \to C} = \frac{1}{2} + \frac{w}{2}\frac{Re_C - Re_D}{e_D - e_C} = \frac{1}{2} - \frac{w}{2}R$ and $p^{C \to D} = \frac{1}{2} + \frac{w}{2}\frac{Re_D - Re_C}{e_D - e_C} = \frac{1}{2} + \frac{w}{2}R$, respectively, and the transition probabilities are $T^{D \to C} = x(1-x)p^{D \to C}$ and $T^{C \to D} = x(1-x)p^{C \to D}$. The evolutionary dynamics of players' strategies under the replicator rule is

$$\frac{dx(t)}{dt} = -w(t)R(t)(1 - x(t))x(t) \qquad (4)$$

We introduce $w(t)$ as a greed parameter that controls the direction and intensity of the evolution of the players' strategy. The greed parameter reflects how much a player prefers to extract more resource by defection than the sustainable level by cooperation. When $w(t) > 0$, the fraction of cooperators $x(t)$ decreases and the fraction of defectors $1 - x(t)$ increases. The larger the absolute value of $w(t)$, the faster the evolution of the players' strategy. When $w(t) < 0$, the opposite occurs. When $w(t) = 0$, the fraction of cooperators remains constant over time.

# Minimal Model

We develop a minimal model to capture the full dynamics of the HES by integrating the evolutionary dynamics of resource and players' strategies. The coupled system is described by Eq. (1). We then perform individual-based simulations of the model, following the microscopic update outlined below. The microscopic update after the initialization consist of the following steps: 1) a player $i$ is randomly chosen; 2) a neighbor of $i$ is randomly chosen, $j$; 3) the strategy of player $i$ to switch to the strategy of player $j$ with probability $p = \frac{1}{2} + \frac{w}{2} \frac{U_j - U_i}{\Delta U_{max}}$; 4) update $x[k] = \frac{N_C}{N}$ where $k$ is the discrete time; then update the resource as

$$R[k] = R[k-1] + \frac{T}{N}\left(R[k-1]\left(1 - \frac{R[k-1]}{K}\right) - R[k-1]\left(x[k-1]\hat{e}_C + (1-x[k-1])\hat{e}_D\right)\right),$$

which is simply the discretization of Eq. (2). The microscopic update reflects the individual-level strategy switching of the players and the subsequent resource update, while the macroscopic ODE captures the population-level dynamics of the fraction of cooperators and resource.

We analyze the stationary solutions of the coupled equations $\{R^*, x^*\}$, which are given by $\mathbf{s}_1 = \{R = 0, \forall x\}$ and $\mathbf{s}_2 = \{R = 1 - \hat{e}_C, x = 1\}$. We then examine the stability of these solutions for the parameter configuration $T, \hat{e}_C, \hat{e}_D, w$ and initial condition $R_0, x_0$. The eigenvalues of the Jacobian matrix at these solutions are: $\boldsymbol{\lambda}_1 = \{0, T(1 - \hat{e}_D(1-x^*) - \hat{e}_C x^*)\}$ and $\boldsymbol{\lambda}_2 = \{T(\hat{e}_C - 1), w(1 - \hat{e}_C)\}$. Depending on the sign of $w$, there are three scenarios.

Scenario $w < 0$: Both $(\boldsymbol{\lambda}_2)_1 < 0$ and $(\boldsymbol{\lambda}_2)_2 < 0$ and $\mathbf{s}_2(R) > 0$, so $\mathbf{s}_2$ is stable and the system sustains the resource. Because $w < 0$, $x(t)$ increases and $x(t) > x_0$. If $x_0 > \frac{-1 + \hat{e}_D}{-\hat{e}_C + \hat{e}_D}$ such that $(\boldsymbol{\lambda}_1)_2 > 0$, $\mathbf{s}_1$ is unstable and the system depletes the resource. Therefore, the system is bi-stable and $x_c^M = \frac{-1 + \hat{e}_D}{-\hat{e}_C + \hat{e}_D}$ is a critical value for ensuring CPR sustainability, i.e., if $x_0 > x_c^M$, the system sustains the resource, regardless of the initial condition of $R_0$.

Scenario $w > 0$: $(\boldsymbol{\lambda}_2)_2 > 0$, so $\mathbf{s}_2$ is unstable. Because $w > 0$, $x(t)$ decreases and $x(t) < x_0$. If $x_0 < \frac{-1 + \hat{e}_D}{-\hat{e}_C + \hat{e}_D}$ such that $(\boldsymbol{\lambda}_1)_2 < 0$, $\mathbf{s}_1$ is stable. Therefore, the system has only one unstable solution that depletes the resource.

Scenario $w = 0$: The evolution of the players' strategies does not change due to $x(t) = x_0$ and the evolution of the resource becomes to $\frac{dR(t)}{dt} = T\left(R(t)\left(1 - \frac{R(t)}{K}\right) - R(t)\left(x_0 \hat{e}_C + (1-x_0)\hat{e}_D\right)\right)$. If $x_0 < x_c^M$, the system has only one unstable solution that depletes the resource $R^* = 0$; if $x_0 > x_c^M$, the system has only one stable solution that sustains the resource $R^* = 1 - \hat{e}_D(1-x^*) - \hat{e}_C x^*$.

## Greed parameter driven by resource availability

We explore a scenario where players adjust their strategies according to the replicator dynamics with a greed parameter that depends on resource availability. To capture this behavior, we assume that $w = f(R)$ where $f()$ is a monotone non-decreasing function with boundary condition $f(0) = -1, f(1) = 1$. We simplify this function to be linear, i.e., $f(R) = aR + b$. To satisfy the boundary condition, we obtain $a = 2, b = -1$. Finally, we integrate the evolutionary dynamics of the resource, Eq. (2), and this new evolutionary dynamics of the players' strategies. The coupled system is

$$\begin{cases} \frac{dR(t)}{dt} = T\left(R(t)\left(1 - \frac{R(t)}{K}\right) - R(t)\left(x(t)\hat{e}_C + (1-x(t))\hat{e}_D\right)\right) \\ \frac{dx(t)}{dt} = -(2R(t)-1)R(t)(1-x(t))x(t) \end{cases} \quad (5)$$

where $K$ can set to 1 by normalizing $R$ for simplicity. We then perform individual-based simulations of the model, following the microscopic rules outlined below. The microscopic rules after the initialization consist of the following steps: 1) a player $i$ is randomly chosen; 2) a neighbor of $i$ is randomly chosen, $j$; 3) the strategy of player $i$ to switch to the strategy of player $j$ with probability $p = \frac{1}{2} + \frac{w}{2}\frac{U_j - U_i}{\Delta U_{max}}$ where $w(t) = 2R(t) - 1$; 4) update $x[k] = \frac{N_C}{N}$ where $k$ is the discrete time; then update the resource as $R[k] = R[k-1] + \frac{T}{N}\left(R[k-1]\left(1 - \frac{R[k-1]}{K}\right) - R[k-1]\left(x[k-1]\hat{e}_C + (1-x[k-1])\hat{e}_D\right)\right)$. The microscopic update reflects the individual-level strategy switching of the players and the subsequent resource update, while the macroscopic ODE captures the population-level dynamics of the fraction of cooperators and resource.

We analyze the stationary solutions of the coupled equations $\{R^*, x^*\}$, which are given by: $\mathbf{s}_1 = \{R = 0, \forall x\}$ where the system depletes the resource, $\mathbf{s}_2 = \{R = \frac{1}{2}, x = \frac{1 - 2\hat{e}_D}{2\hat{e}_C - 2\hat{e}_D}\}$ for $0 < \hat{e}_C < 1/2$ where the system sustains the resource and $s_3 = \{R = 1 - \hat{e}_C, x = 1\}$ where the system sustains the resource. The stability of these solutions cannot be easily assessed by the sign of the eigenvalue, but we can estimate it by the determinant and trace of the two-dimensional Jacobian matrix. If the determinant is positive and the trace is negative, i.e., $\text{Det} > 0 \wedge \text{Tr} < 0$, the solution is stable. Using this method, we determine that $s_1$ is neutral stable for $x(t) < \frac{-1 + \hat{e}_D}{-\hat{e}_C + \hat{e}_D}$, $s_2$ is stable for $\hat{e}_C < 1/2$ and $s_3$ is stable for $\hat{e}_C > 1/2$. As $R_0 \to 0^+$, $w = f(R) \to -1^+$ so that $x(t)$ increases more rapidly. Therefore, $x_c^R = x_c^M = \frac{-1 + \hat{e}_D}{-\hat{e}_C + \hat{e}_D}$ is a critical value for ensuring CPR sustainability for any initial condition $\{R_0, x_0\}$, i.e., if initial condition $x_0 > x_c^R$, then the system sustains the resource; otherwise, it may deplete the resource.

## Greed parameter driven by conformity effect

We explore a scenario where players adjust their strategies according to the replicator dynamics with a greed parameter that reflects conformity effect. To capture this behavior, we assume that $w = f(R)$ where $f()$ is a monotone non-decreasing function with boundary condition $f(1) = -1, f(0) = 1$. We simplify this function to be linear, i.e., $f(x) = ax + b$. To satisfy the boundary condition, we obtain $a = -2, b = 1$. Finally, we integrate the evolutionary dynamics of the resource, Eq. (2), and this new evolutionary dynamics of the players' strategies. The coupled system is

$$\begin{cases} \dfrac{dR(t)}{dt} = T\left(R(t)\left(1 - \dfrac{R(t)}{K}\right) - R(t)\left(x(t)\hat{e}_C + (1 - x(t))\hat{e}_D\right)\right) \\ \dfrac{dx(t)}{dt} = -(1 - 2x(t))R(t)(1 - x(t))x(t) \end{cases} \quad (6)$$

where $K$ can set to 1 by normalizing $R$ for simplicity. We then perform individual-based simulations of the model, following the microscopic rules outlined below. The microscopic rules after the initialization consist of the following steps: 1) a player $i$ is randomly chosen; 2) a neighbor of $i$ is randomly chosen, $j$; 3) the strategy of player $i$ to switch to the strategy of player $j$ with probability $p = \dfrac{1}{2} + \dfrac{w}{2}\dfrac{U_j - U_i}{\Delta U_{max}}$ where $w(t) = 1 - 2x(t)$; 4) update $x[k] = \dfrac{N_C}{N}$ where $k$ is the discrete time; then update the resource as $R[k] = R[k-1] + \dfrac{T}{N}\left(R[k-1]\left(1 - \dfrac{R[k-1]}{K}\right) - R[k-1]\left(x[k-1]\hat{e}_C + (1 - x[k-1])\hat{e}_D\right)\right)$. The microscopic update reflects the individual-level strategy switching of the players and the subsequent resource update, while the macroscopic ODE captures the population-level dynamics of the cooperators fraction and resource.

We analyze the stationary solutions of the coupled equations $\{R^*, x^*\}$, which are given by: $\mathbf{s}_1 = \{R = 0, \forall x\}$ where the system depletes the resource, $\mathbf{s}_2 = \{R = 1 - \hat{e}_C, x = 1\}$ where the system sustains the resource and $\mathbf{s}_3 = \{R = 1 - \frac{\hat{e}_C}{2} - \frac{\hat{e}_D}{2}, x = \frac{1}{2}\}$ for $\hat{e}_D < 2 - \hat{e}_C$ where the system sustains the resource. Using method of $\mathrm{Det} > 0 \wedge \mathrm{Tr} < 0$, we determine that $\mathbf{s}_1$ is neutral stable for $x(t) < x_c^M$, $\mathbf{s}_2$ is stable and $\mathbf{s}_3$ is unstable. When $x(t) < 1/2$ such that $w(t) = 1 - 2x(t) > 0$, $x(t)$ decreases and $\mathbf{s}_1$ is satisfied. When $x(t) > 1/2$ such that $w(t) = 1 - 2x(t) < 0$, $x(t)$ increases and $\mathbf{s}_1$ becomes unstable due to $x(t) > x_c^M$. Moreover, when $x_0 > x_c^M$, $\mathbf{s}_2$ is the only stable solution. Therefore, $x_c^C = \max\{x_c^M, 1/2\}$ is a critical value for ensuring CPR sustainability for any initial condition $\{R_0, x_0\}$, i.e., if initial condition $x_0 > x_c^C$, then the system sustains the resource; otherwise, it may deplete the resource.

## Greed parameter driven by both resource availability and conformity effect

We explore a scenario where players adjust their strategies according to the replicator dynamics with a greed parameter that depends on both resource availability and conformity effect. To capture this behavior, we assume that $w = f(R, x)$ where $f()$ is a monotone non-decreasing function for $R$ and a monotone non-increasing function for $x$ with boundary condition $f(0,1) = -1, f(1,0) = 1$. We simplify this function to be linear, i.e., $f(R, x) = aR + bx + c$. To satisfy the boundary condition, we obtain $a = 1 - c, b = -1 - c$ where $-1 \leq c \leq 1$. Finally, we integrate the evolutionary dynamics of the resource, Eq. (2), and this new evolutionary dynamics of the players' strategies. The coupled system is

$$\begin{cases} \dfrac{dR(t)}{dt} = T\left[R(t)\left(1 - \dfrac{R(t)}{K}\right) - R(t)\left(x(t)\hat{e}_C + (1 - x(t))\hat{e}_D\right)\right] \\ \dfrac{dx(t)}{dt} = -\left((1-c)R(t) + (-1-c)x(t) + c\right)R(t)(1 - x(t))x(t) \end{cases} \quad (7)$$

where $K$ can set to 1 by normalizing $R$ for simplicity. We then perform individual-based simulations of the model, following the microscopic rules outlined below. The microscopic rules after the initialization consist of the following steps: 1) a player $i$ is randomly chosen; 2) a neighbor of $i$ is randomly chosen, $j$; 3) the strategy of player $i$ to switch to the strategy of player $j$ with probability $p = \dfrac{1}{2} + \dfrac{w}{2} \dfrac{U_j - U_i}{\Delta U_{max}}$ where $w = (1-c)R(t) + (-1-c)x(t) + c$; 4) update $x[k] = \dfrac{N_C}{N}$ where $k$ is the discrete time; then update the resource as

$$R[k] = R[k-1] + \frac{T}{N}\left(R[k-1]\left(1 - \frac{R[k-1]}{K}\right) - R[k-1]\left(x[k-1]\hat{e}_C + (1-x[k-1])\hat{e}_D\right)\right).$$ The microscopic update reflects the individual-level strategy switching of the players and the subsequent resource update, while the macroscopic ODE captures the population-level dynamics of the fraction of cooperators and resource.

We analyze the stationary solutions of the coupled equations $\{R^*, x^*\}$, which are given by: $\mathbf{s}_1 = \{R = 0, \forall x\}$ where the system depletes the resource, $\mathbf{s}_2 = \{R = 1 - \hat{e}_C, x = 1\}$ where the system sustains the resource and $\mathbf{s}_3 = \{R = \frac{1 + c - \hat{e}_D - c\hat{e}_D}{1 + \hat{e}_C - \hat{e}_D + c(1 - \hat{e}_C + \hat{e}_D)}, x = \frac{1 - \hat{e}_D + c\hat{e}_D}{1 + \hat{e}_C - \hat{e}_D + c(1 - \hat{e}_C + \hat{e}_D)}\}$ where the system sustains the resource. Using method of $\text{Det} > 0 \wedge \text{Tr} < 0$, we determine that $\mathbf{s}_1$ is neutral stable for $x(t) < x_c^M$, $\mathbf{s}_2$ is stable and $\mathbf{s}_3$ is stable for $0 < \hat{e}_C < 1/2$, $-1 < c < \frac{\hat{e}_C}{\hat{e}_C - 1}$ and $T > -\frac{(c+1)(c(\hat{e}_C - 1) - \hat{e}_C)((c-1)\hat{e}_D + 1)}{(c(-\hat{e}_C) + (c-1)\hat{e}_D + c + \hat{e}_C + 1)^2}$. This new $w = (1-c)R(t) + (-1-c)x(t) + c$ represents a balance between the resource availability and conformity effect with weights $1-c$ and $-1-c$, respectively. Its critical value is a weighted average of their critical values, i.e., $(1-c)x_c^R$ for the resource-driven greed parameter and $(-1-c)x_c^C$ for the conformity-driven greed parameter. These two components need to be rescaled to satisfy the boundary condition $0 \leq x \leq 1$: When $c \to 1$, the critical value approximates that of the resource-driven greed parameter. When $c \to -1$, the critical value approximates that of the conformity-driven greed parameter. Therefore, $x_c^{RC} = \frac{1-c}{2}x_c^R + \frac{-1-c}{2}x_c^C = \frac{1-c}{2}x_c^M + \frac{-1-c}{2}\max\{x_c^M, 1/2\}$ is a critical value for ensuring CPR sustainability for any initial condition $\{R_0, x_0\}$, i.e., if initial condition $x_0 > x_c^{RC}$, then the system sustains the resource; otherwise, it may deplete the resource.

# Code availability

Code for all analyses is available at OSF (https://osf.io/s7j3h/).

# Acknowledgments

This work was supported by Microsoft AI for Earth, Experimental Social Science Laboratory at University of California Berkeley, Natural Science Fund of Zhejiang Province under Grant No. LZ18G010001, LZ22G010001, Science Foundation of Zhejiang

Sci-Tech University under Grant No. 18092125-Y, 22092034-Y.

# Author Contributions Statement

Conceptualization: C.T. Methodology: C.T. Investigation: C.T. Visualization: C.T. Funding acquisition: C.T., R.C., Y.F., X.P. Project administration: C.T. Writing, original draft: C.T. Writing, review & editing: C.T., R.C., Y.F., X.P.

# Competing Interests Statement

Authors declare that they have no competing interests.

# Supplementary Notes

## Continuous approximation of strategies evolutionary dynamics

The players' strategies evolve according to the standard replicator rule, whereby a player switches strategy depending on the payoff differences with one of the neighbors. The transition probability at time $t$ from $N_C$ cooperators to $N_C \pm 1$, given a resource level $R(t)$ and a greed parameter $w(t)$, is given by

$$T^{D \to C}(N_C | R, w; t) = p^{D \to C} \frac{N_C}{N} \frac{N - N_C}{N}, \quad T^{C \to D}(N_C | R, w; t) = p^{C \to D} \frac{N_C}{N} \frac{N - N_C}{N}$$

where $p^{D \to C} = \frac{1}{2} + \frac{w}{2} \frac{Re_C - Re_D}{e_D - e_C} = \frac{1}{2} - \frac{w}{2} R$ and $p^{C \to D} = \frac{1}{2} + \frac{w}{2} \frac{Re_D - Re_C}{e_D - e_C} = \frac{1}{2} + \frac{w}{2} R$ are the probabilities of switching strategy from defection to cooperation, and vice versa, respectively, with $-1 \leq w \leq 1$. If players interact through a complete network, i.e., each node is connected to all other nodes, then the evolution of the probability $P^\tau(N_C)$ of having $N_C$ cooperators at time $\tau$ is governed by the Master Equation

$$P^{\tau+1}(N_C) - P^\tau(N_C) = P^\tau(N_C - 1) T^{D \to C}(N_C - 1 | R, w; \tau) + P^\tau(N_C + 1) T^{C \to D}(N_C + 1 | R, w; \tau)$$
$$- P^\tau(N_C) T^{C \to D}(N_C | R, w; \tau) - P^\tau(N_C) T^{D \to C}(N_C | R, w; \tau)$$

Using a classic evolutionary game framework, we can derive a general Fokker-Planck equation for the probability density of having the fraction of cooperators $x$, $\rho(x)$. We start from the Master Equation and substitute $x = \frac{N_C}{N}, t = \frac{\tau}{N}$ and $\rho(x; t) = N P^\tau(N_C)$ to obtain

$$\rho(x; t + N^{-1}) - \rho(x; t) = \rho(x - N^{-1}; t) T^{D \to C}(x - N^{-1} | R, w; t) + \rho(x + N^{-1}; t) T^{C \to D}(x + N^{-1} | R, w; t)$$
$$- \rho(x; t) T^{C \to D}(x | R, w; t) - \rho(x; t) T^{D \to C}(x | R, w; t)$$

For $N$ much larger than 1, we can approximate the probability densities and the transition probabilities by a Taylor series expansion around $x(t)$. By neglecting higher order terms in $N^{-1}$, we obtain the Fokker-Planck equation

$$\frac{d}{dt} \rho(x; t) = -\frac{d}{dx}[a(x | R, w; t) \rho(x; t)] + \frac{1}{2} \frac{d^2}{dx^2}[b^2(x | R, w; t) \rho(x; t)]$$

where $a(x | R, w; t) = T^+(x | R, w; t) - T^-(x | R, w; t)$ and $b(x | R, w; t) = \sqrt{\frac{1}{N}(T^+(x | R, w; t) + T^-(x | R, w; t))}$.

Since the time steps are independent, the noise is uncorrelated in time and we can apply the Ito calculus to derive the corresponding Langevin equation:

$$\frac{dx(t)}{dt} = a(x \mid R, w; t) + b(x \mid R, w; t)\zeta$$

where $\zeta$ is an uncorrelated Gaussian white noise. For $N \to \infty$, the diffusion term $b(x \mid R, w; t)$ disappears with $1/\sqrt{N}$ and a deterministic equation is obtained. Therefore, the evolutionary dynamics of players' strategies under the replicator rule is

$$\frac{dx(t)}{dt} = T^{D \to C} - T^{C \to D}$$

# Extension of greed parameter driven by resource availability

In the main text, we assume that the function of the greed parameter driven by resource availability is linear form, i.e., $w = f(R) = aR + b$. Here, we extend our analysis to a quadratic form, i.e., $w = f(R) = aR^2 + bR + c$. To ensure $f()$ is a monotone non-decreasing function with boundary condition $f(0) = -1, f(1) = 1$, we obtain $a + b = 2, c = -1$ where $-2 \leq a \leq 2$. The resulting coupled system is then given by

$$\begin{cases} \dfrac{dR(t)}{dt} = T\left(R(t)\left(1 - \dfrac{R(t)}{K}\right) - R(t)\left(x(t)\hat{e}_C + (1 - x(t))\hat{e}_D\right)\right) \\ \dfrac{dx(t)}{dt} = -\left(aR(t)^2 + bR(t) - 1\right)R(t)(1 - x(t))x(t) \end{cases} \quad (1)$$

where $K$ can set to 1 by normalizing $R$ for simplicity and other parameters are the same as in the main text.

The properties and stability of the solution are difficult to analyze analytically, but we use numerical simulations to explore the dynamics of the system. We find that they exhibit a similar pattern as in the linear case, where the system shows different behaviors depending on the parameter configuration and initial condition (see Supplementary Fig. 19). Additionally, the critical value under the linear form $x_c^R = x_c^M = \dfrac{-1 + \hat{e}_D}{-\hat{e}_C + \hat{e}_D}$ still works well for ensuring the CPR sustainability under the quadratic form.

# Extension of greed parameter driven by conformity effect

In the main text, we assume that the function of the greed parameter driven by conformity effect is linear form, i.e., $w = f(x) = ax + b$. Here, we extend our analysis to a quadratic form, i.e., $w = f(x) = ax^2 + bx + c$. To ensure $f()$ is a monotone non-decreasing function with boundary condition $f(1) = -1, f(0) = 1$, we obtain $a + b = -2, c = 1$ where

$-2 \leq a \leq 2$. The resulting coupled system is then given by

$$\begin{cases} \dfrac{dR(t)}{dt} = T\left(R(t)\left(1-\dfrac{R(t)}{K}\right) - R(t)\left(x(t)\hat{e}_C + (1-x(t))\hat{e}_D\right)\right) \\ \dfrac{dx(t)}{dt} = -\left(ax(t)^2 + bx(t) + 1\right)R(t)(1-x(t))x(t) \end{cases} \quad (2)$$

where $K$ can set to 1 by normalizing $R$ for simplicity and other parameters are the same as in the main text.

The properties and stability of the solution are difficult to analyze analytically, but we use numerical simulations to explore the dynamics of the system. We find that they exhibit a similar pattern as in the linear case, where the system shows different behaviors depending on the parameter configuration and initial condition (see Supplementary Fig. 20). The critical value of the conformity-driven model is the maximum value of the feasible solution of $f(x)=0$ and $x_c^M$. Under the quadratic form, the feasible solution is given by $\dfrac{-2-a+\sqrt{4+a^2}}{2a}$, which is different from the linear form, where the feasible solution is 0.5. Therefore, under the quadratic form, the critical value of the conformity-driven model is

$$x_c^C = \max\{x_c^M, \dfrac{-2-a+\sqrt{4+a^2}}{2a}\}.$$

# Extension of greed parameter driven by both resource availability and conformity effect

In the main text, we assume that the function of the greed parameter driven by both resource availability and conformity effect is linear form, i.e., $w = f(R,x) = aR + bx + c$. Here, we extend our analysis to a quadratic form, i.e., $w = f(x) = aR^2 + bR + cx^2 + dx + e$. To ensure $f()$ is a monotone non-decreasing function for $R$ and a monotone non-increasing function for $x$ with boundary condition $f(0,1) = -1, f(1,0) = 1$, we obtain $c+d+e = -1, a+b+e = 1$. The resulting coupled system is then given by

$$\begin{cases} \dfrac{dR(t)}{dt} = T\left(R(t)\left(1-\dfrac{R(t)}{K}\right) - R(t)\left(x(t)\hat{e}_C + (1-x(t))\hat{e}_D\right)\right) \\ \dfrac{dx(t)}{dt} = -\left(aR(t)^2 + bR(t) + cx(t)^2 + dx(t) + e\right)R(t)(1-x(t))x(t) \end{cases} \quad (3)$$

where $K$ can set to 1 by normalizing $R$ for simplicity and other parameters are the same as in the main text.

The properties and stability of the solution are difficult to analyze analytically, but we use numerical simulations to explore the dynamics of the system. We find that they exhibit a similar pattern as in the linear case, where the system shows different behaviors depending on the parameter configuration and initial condition (see Supplementary Fig. 21). In this case, we cannot estimate the critical value analytically, because the quadratic form involves too many variables and complexity.

## Effect of the social network structure

So far, we have assumed that all players can interact with all the other players in the game, without accounting for the effect of the social network structure among players. However, in reality, players do not have access to all the other players; rather they are only connected with a few other players and the network through which players interact has a given structure. To test the effect of such a network structure on the CPR sustainability, we perform numerical simulations, where players compare their payoff only with players they are connected to through given social networks. In other words, we simulate spatially explicit dynamics. If the number of players is large enough, stochasticity may be so small that it cannot drive the system far from the expected stable equilibrium obtained by the mean-field approach. We compare the outcomes of microscopic update and macroscopic ODE with parameter configuration $\hat{e}_C = 0.7, \hat{e}_D = 1.1$ for all models under two network structures: Barabasi-Albert (BA) and small-world (SW) networks (see Supplementary Fig. 22-24).

# Supplementary Figures

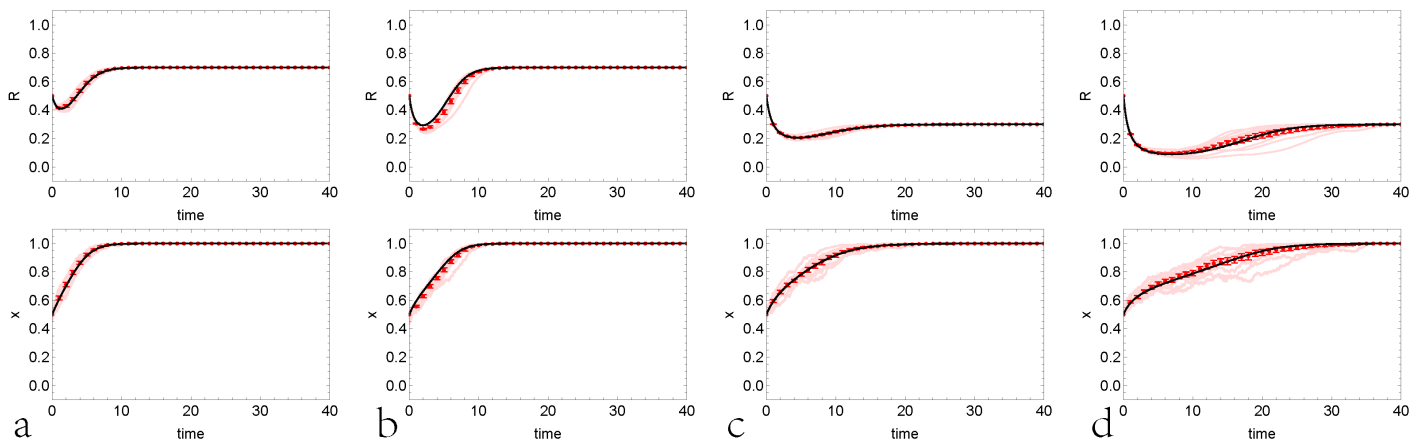

**Supplementary Fig. 1 | Comparison between macroscopic ODE and microscopic update for the minimal model with** $w = -1$. **(a)** $\hat{e}_C = 0.3, \hat{e}_D = 1.1$; **(b)** $\hat{e}_C = 0.3, \hat{e}_D = 1.5$; **(c)** $\hat{e}_C = 0.7, \hat{e}_D = 1.1$; **(d)** $\hat{e}_C = 0.7, \hat{e}_D = 1.5$. The other parameters are growth rate $T = 2$, player number $N = 500$ and initial condition $R_0 = 0.5, x_0 = 0.5$ as well as independent realization number $n = 10$. Each light red line is one realization of numerical simulation and the black line is the theoretical prediction of macroscopic ODE. The red points and red fences are the average and stand error of all realizations at each integer time.

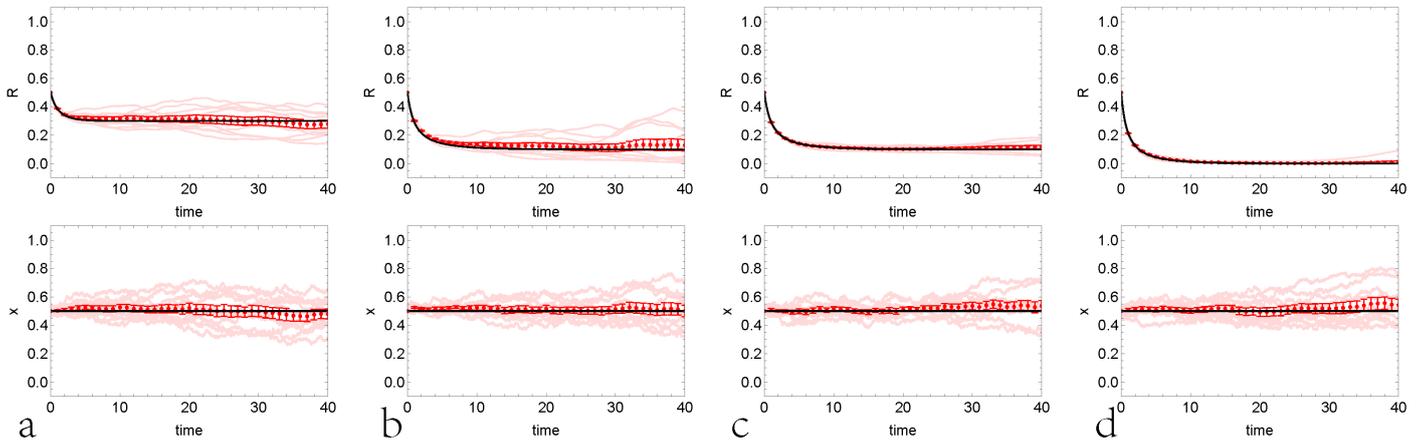

**Supplementary Fig. 2 | Comparison between macroscopic ODE and microscopic update for the minimal model with** $w=0$. **(a)** $\hat{e}_C = 0.3, \hat{e}_D = 1.1$; **(b)** $\hat{e}_C = 0.3, \hat{e}_D = 1.5$; **(c)** $\hat{e}_C = 0.7, \hat{e}_D = 1.1$; **(d)** $\hat{e}_C = 0.7, \hat{e}_D = 1.5$. The other parameters are growth rate $T=2$, player number $N=500$ and initial condition $R_0 = 0.5, x_0 = 0.5$ as well as independent realization number $n=10$. Each light red line is one realization of numerical simulation and the black line is the theoretical prediction of macroscopic ODE. The red points and red fences are the average and stand error of all realizations at each integer time.

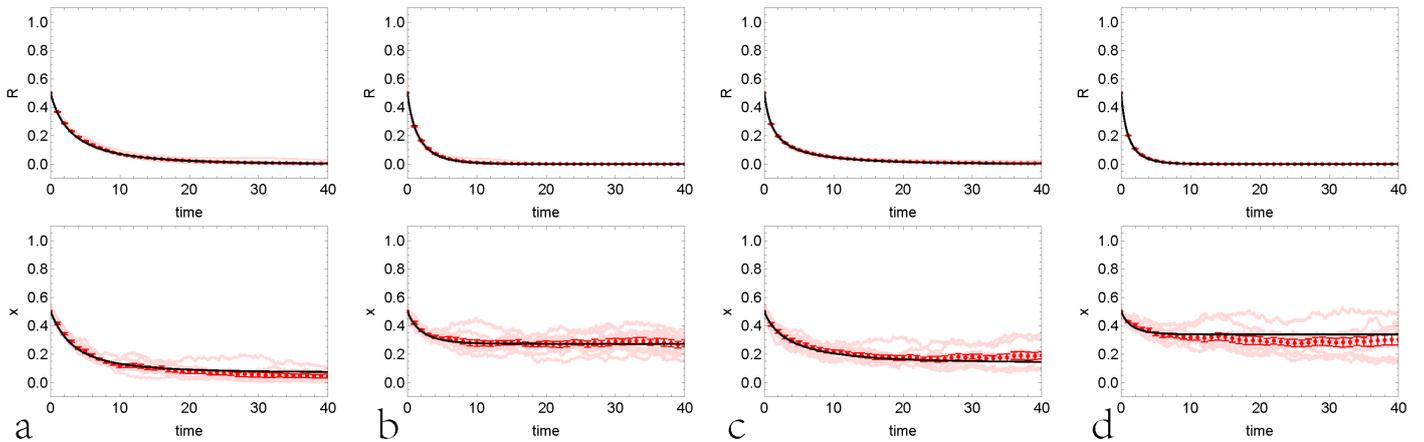

**Supplementary Fig. 3 | Comparison between macroscopic ODE and microscopic update for the minimal model with** $w=1$. **(a)** $\hat{e}_C = 0.3, \hat{e}_D = 1.1$; **(b)** $\hat{e}_C = 0.3, \hat{e}_D = 1.5$; **(c)** $\hat{e}_C = 0.7, \hat{e}_D = 1.1$; **(d)** $\hat{e}_C = 0.7, \hat{e}_D = 1.5$. The other parameters are growth rate $T=2$, player number $N=500$ and initial condition $R_0 = 0.5, x_0 = 0.5$ as well as independent realization number $n=10$. Each light red line is one realization of numerical simulation and the black line is the theoretical prediction of macroscopic ODE. The red points and red fences are the average and stand error of all realizations at each integer time.

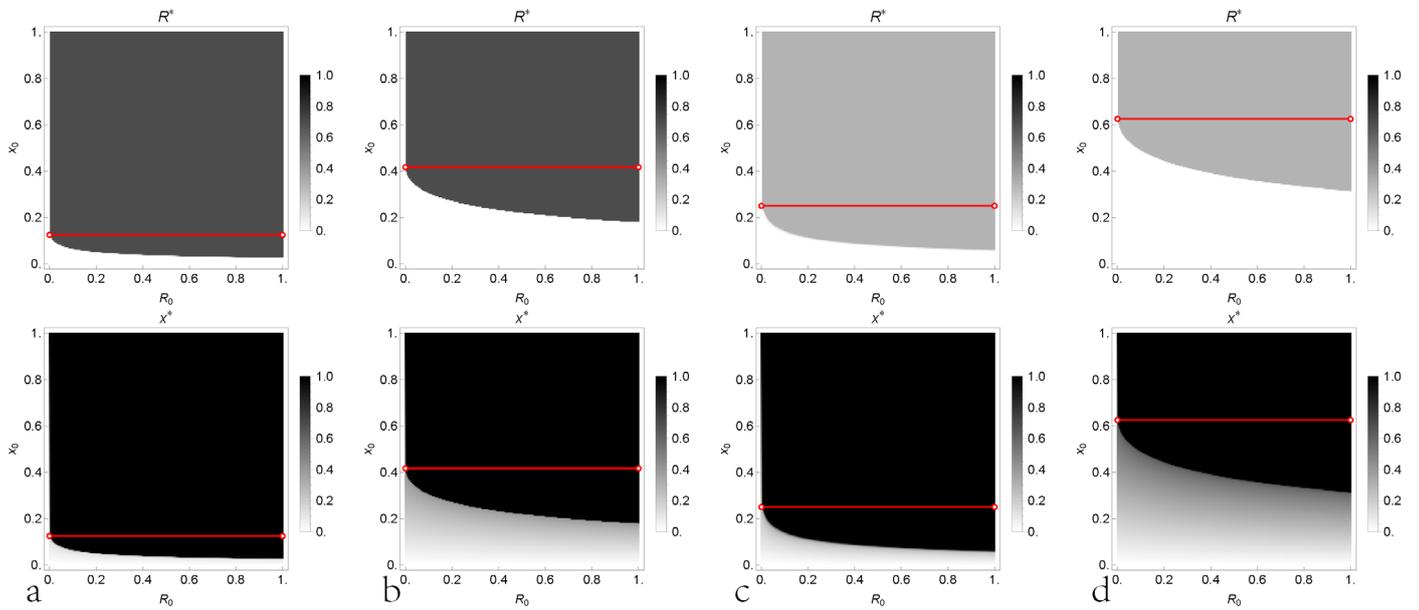

**Supplementary Fig. 4 |** Density plot of stable equilibrium $R^*$ and $x^*$ for the minimal model with $w=-1$. **(a)** $\hat{e}_C=0.3, \hat{e}_D=1.1$; **(b)** $\hat{e}_C=0.3, \hat{e}_D=1.5$; **(c)** $\hat{e}_C=0.7, \hat{e}_D=1.1$; **(d)** $\hat{e}_C=0.7, \hat{e}_D=1.5$. The other parameters are growth rate $T=2$.

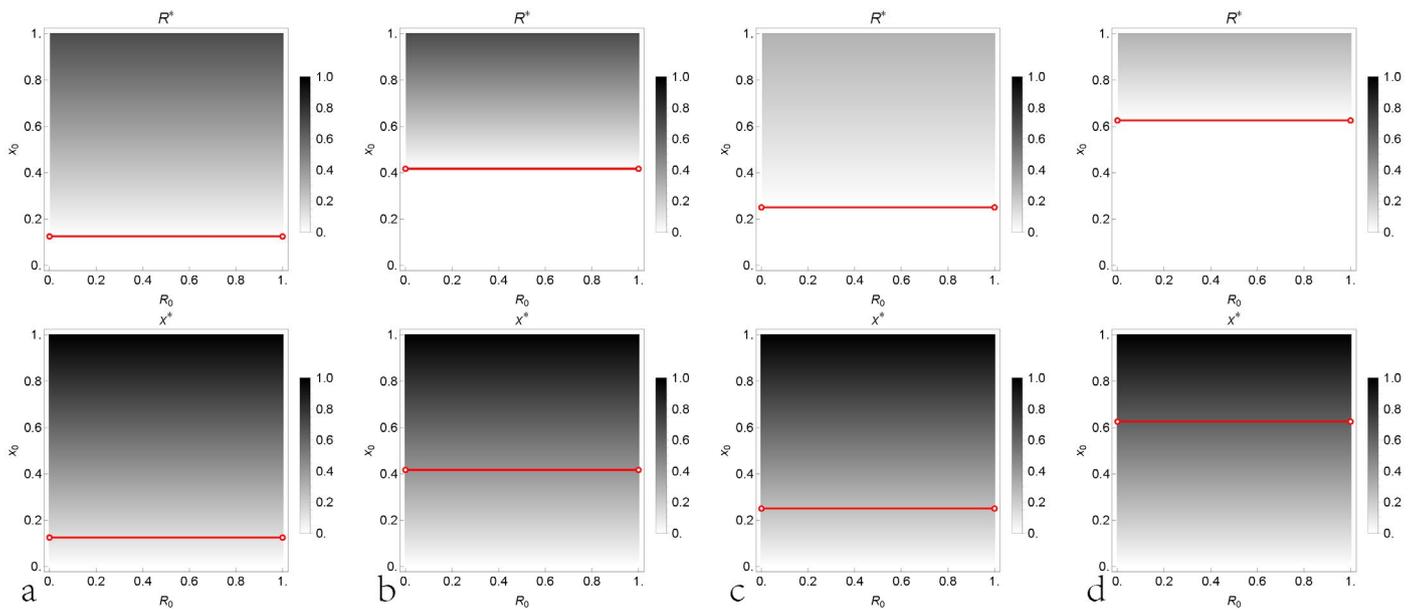

**Supplementary Fig. 5 |** Density plot of stable equilibrium $R^*$ and $x^*$ for the minimal model with $w=0$. **(a)** $\hat{e}_C=0.3, \hat{e}_D=1.1$; **(b)** $\hat{e}_C=0.3, \hat{e}_D=1.5$; **(c)** $\hat{e}_C=0.7, \hat{e}_D=1.1$; **(d)** $\hat{e}_C=0.7, \hat{e}_D=1.5$. The other parameters are growth rate $T=2$.

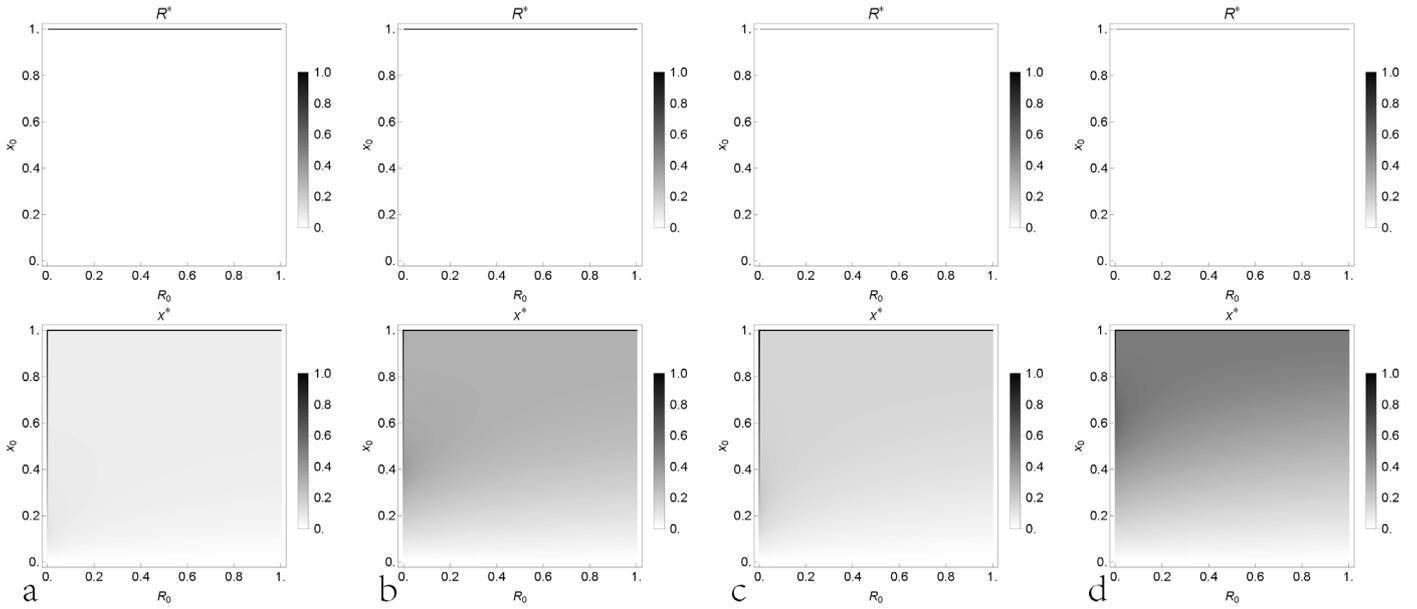

**Supplementary Fig. 6 | Density plot of stable equilibrium $R^*$ and $x^*$ for the minimal model with $w=1$.** (a) $\hat{e}_C = 0.3, \hat{e}_D = 1.1$; (b) $\hat{e}_C = 0.3, \hat{e}_D = 1.5$; (c) $\hat{e}_C = 0.7, \hat{e}_D = 1.1$; (d) $\hat{e}_C = 0.7, \hat{e}_D = 1.5$. The other parameters are growth rate $T=2$.

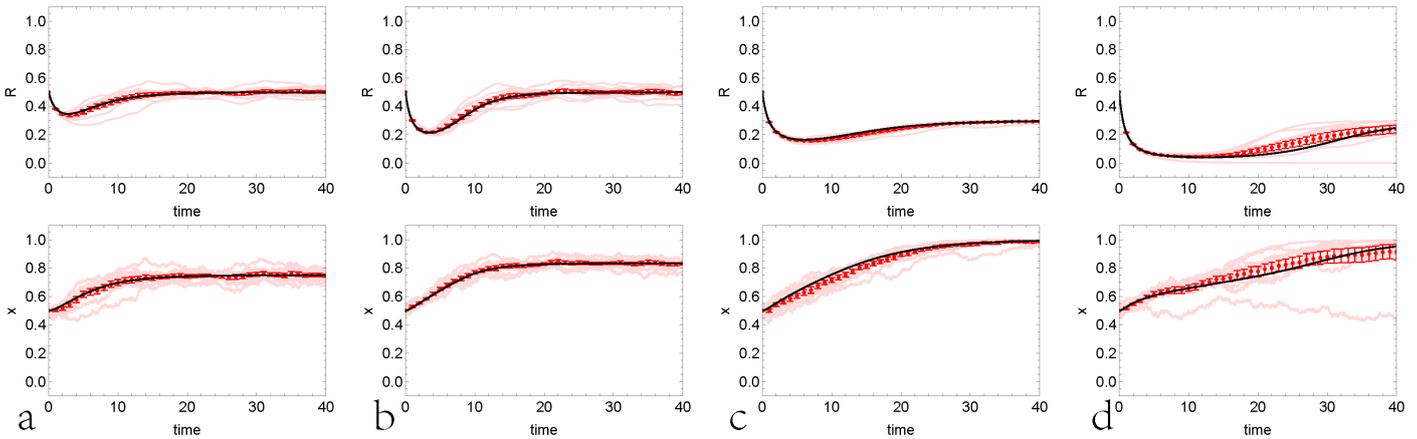

**Supplementary Figure 7 | Comparison between macroscopic ODE and microscopic update for the resource-driven model.** (a) $\hat{e}_C = 0.3, \hat{e}_D = 1.1$; (b) $\hat{e}_C = 0.3, \hat{e}_D = 1.5$; (c) $\hat{e}_C = 0.7, \hat{e}_D = 1.1$; (d) $\hat{e}_C = 0.7, \hat{e}_D = 1.5$. The other parameters are growth rate $T=2$, player number $N=500$ and initial condition $R_0 = 0.5, x_0 = 0.5$ as well as independent realization number $n=10$. Each light red line is one realization of numerical simulation and the black line is the theoretical prediction of macroscopic ODE. The red points and red fences are the average and stand error of all realizations at each integer time.

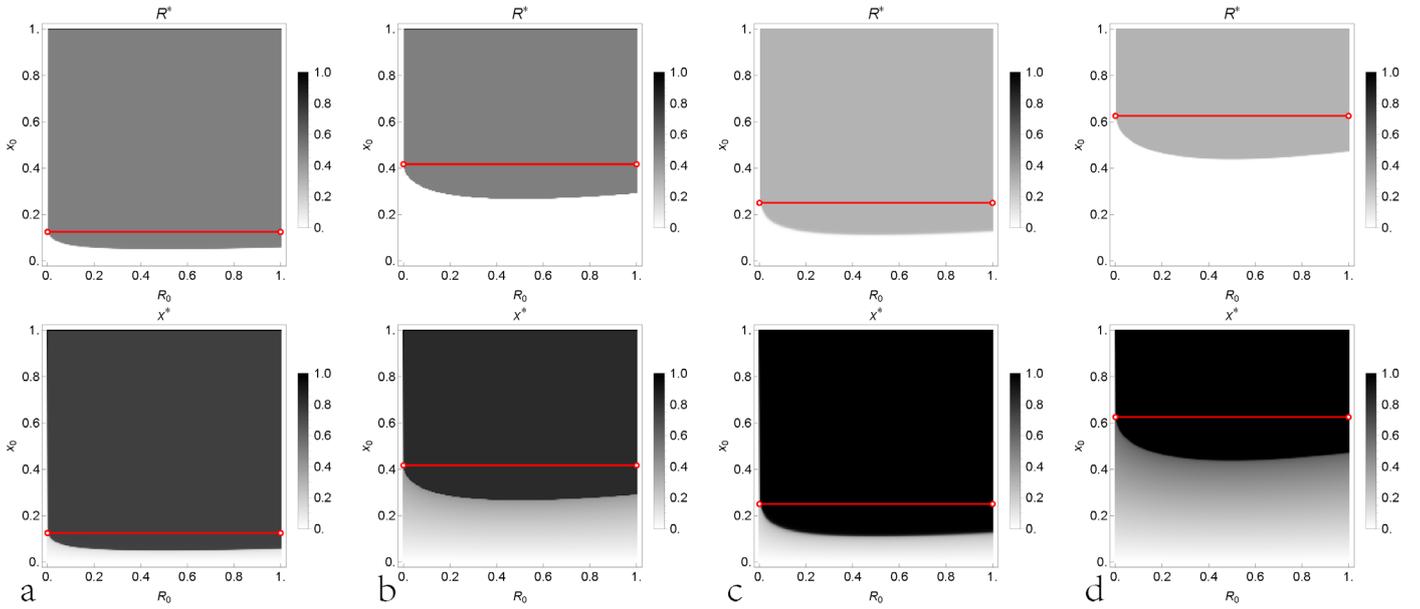

**Supplementary Fig. 8 | Density plot of stable equilibrium $R^*$ and $x^*$ for the resource-driven model.** (a) $\hat{e}_C = 0.3, \hat{e}_D = 1.1$; (b) $\hat{e}_C = 0.3, \hat{e}_D = 1.5$; (c) $\hat{e}_C = 0.7, \hat{e}_D = 1.1$; (d) $\hat{e}_C = 0.7, \hat{e}_D = 1.5$. The other parameters are growth rate $T = 2$.

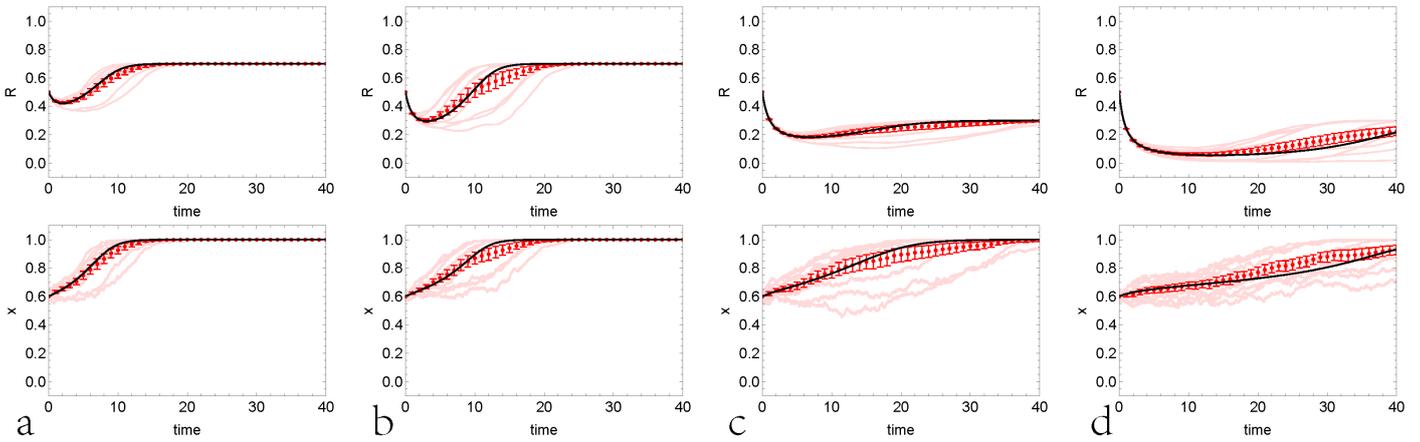

**Supplementary Figure 9 | Comparison between macroscopic ODE and microscopic update for the conformity-driven model.** (a) $\hat{e}_C = 0.3, \hat{e}_D = 1.1$; (b) $\hat{e}_C = 0.3, \hat{e}_D = 1.5$; (c) $\hat{e}_C = 0.7, \hat{e}_D = 1.1$; (d) $\hat{e}_C = 0.7, \hat{e}_D = 1.5$. The other parameters are growth rate $T = 2$, player number $N = 500$ and initial condition $R_0 = 0.5, x_0 = 0.5$ as well as independent realization number $n = 10$. Each light red line is one realization of numerical simulation and the black line is the theoretical prediction of macroscopic ODE. The red points and red fences are the average and stand error of all realizations at each integer time.

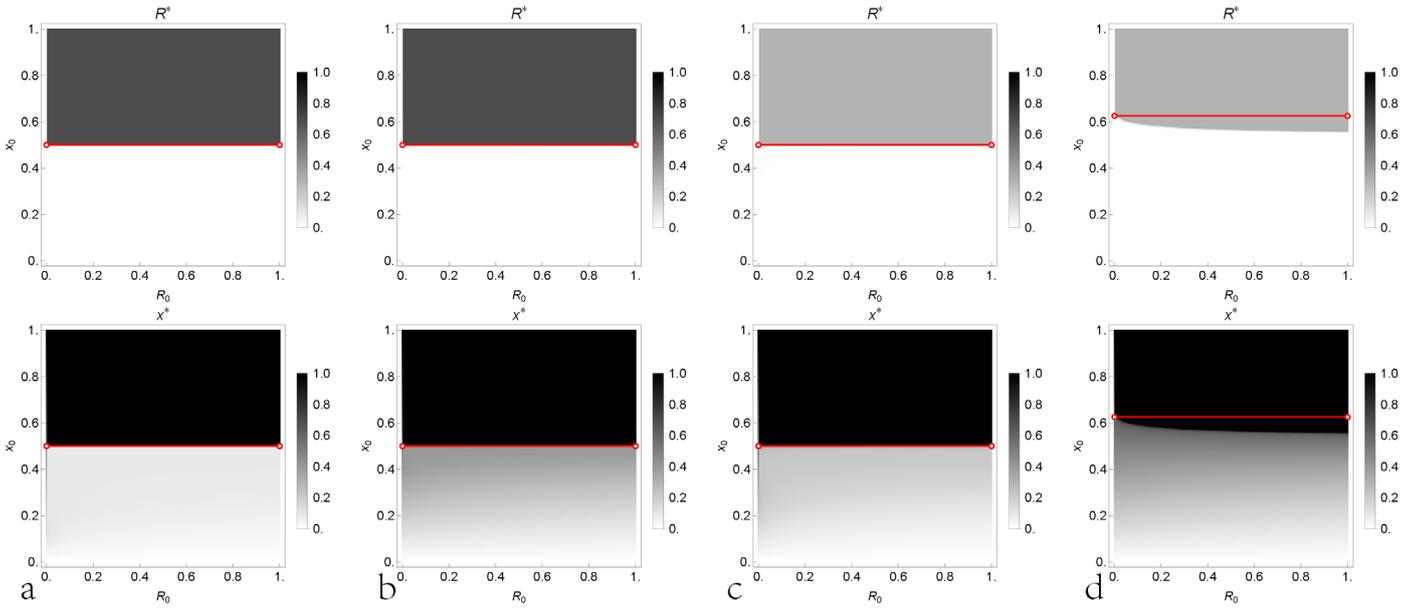

**Supplementary Fig. 10 | Density plot of stable equilibrium $R^*$ and $x^*$ for the conformity-driven model.** (a) $\hat{e}_C = 0.3, \hat{e}_D = 1.1$; (b) $\hat{e}_C = 0.3, \hat{e}_D = 1.5$; (c) $\hat{e}_C = 0.7, \hat{e}_D = 1.1$; (d) $\hat{e}_C = 0.7, \hat{e}_D = 1.5$. The other parameters are growth rate $T = 2$.

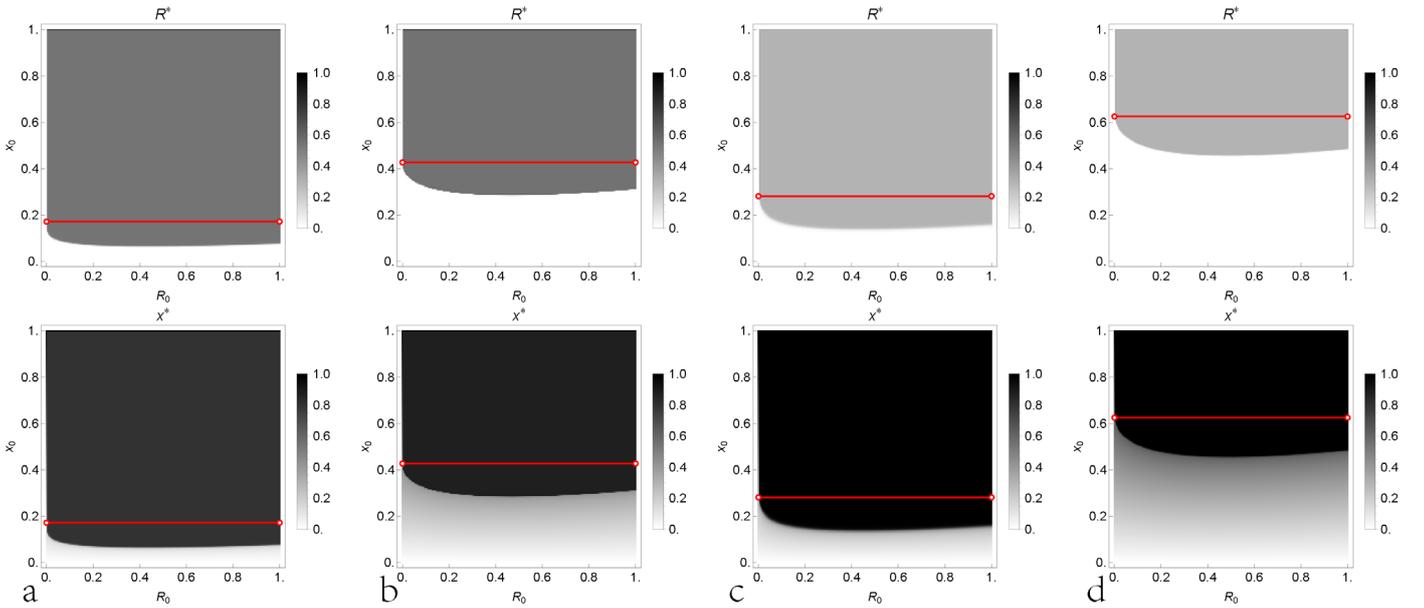

**Supplementary Figure 11 | Comparison between macroscopic ODE and microscopic update for the resource and conformity-driven model with $c = -0.75$.** (a) $\hat{e}_C = 0.3, \hat{e}_D = 1.1$; (b) $\hat{e}_C = 0.3, \hat{e}_D = 1.5$; (c) $\hat{e}_C = 0.7, \hat{e}_D = 1.1$; (d) $\hat{e}_C = 0.7, \hat{e}_D = 1.5$. The other parameters are growth rate $T = 2$, player number $N = 500$ and initial condition $R_0 = 0.5, x_0 = 0.5$ as well as independent realization number $n = 10$. Each light red line is one realization of numerical simulation and the black line is the theoretical prediction of macroscopic ODE. The red points and red fences are the average and stand error of all realizations at each integer time.

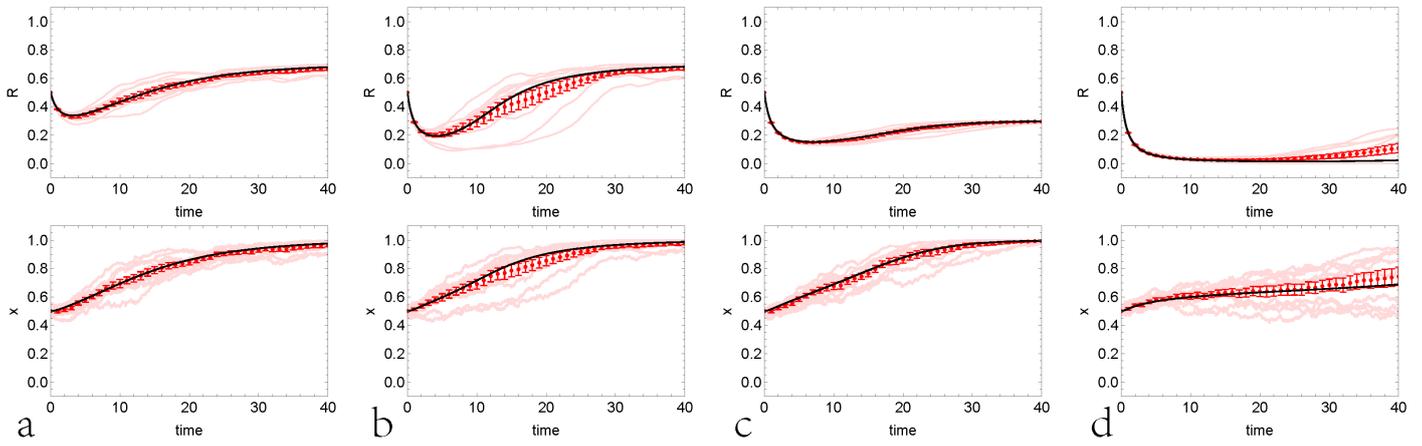

**Supplementary Figure 12 | Comparison between macroscopic ODE and microscopic update for the resource and conformity-driven model with** $c = -0.25$. **(a)** $\hat{e}_C = 0.3, \hat{e}_D = 1.1$; **(b)** $\hat{e}_C = 0.3, \hat{e}_D = 1.5$; **(c)** $\hat{e}_C = 0.7, \hat{e}_D = 1.1$; **(d)** $\hat{e}_C = 0.7, \hat{e}_D = 1.5$. The other parameters are growth rate $T = 2$, player number $N = 500$ and initial condition $R_0 = 0.5, x_0 = 0.5$ as well as independent realization number $n = 10$. Each light red line is one realization of numerical simulation and the black line is the theoretical prediction of macroscopic ODE. The red points and red fences are the average and stand error of all realizations at each integer time.

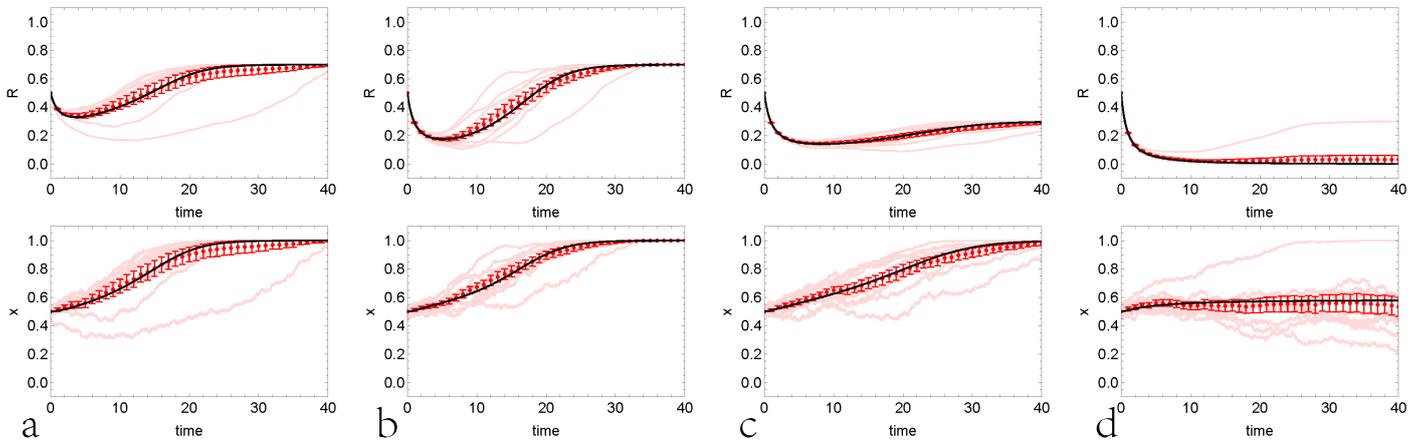

**Supplementary Figure 13 | Comparison between macroscopic ODE and microscopic update for the resource and conformity-driven model with** $c = 0.25$. **(a)** $\hat{e}_C = 0.3, \hat{e}_D = 1.1$; **(b)** $\hat{e}_C = 0.3, \hat{e}_D = 1.5$; **(c)** $\hat{e}_C = 0.7, \hat{e}_D = 1.1$; **(d)** $\hat{e}_C = 0.7, \hat{e}_D = 1.5$. The other parameters are growth rate $T = 2$, player number $N = 500$ and initial condition $R_0 = 0.5, x_0 = 0.5$ as well as independent realization number $n = 10$. Each light red line is one realization of numerical simulation and the black line is the theoretical prediction of macroscopic ODE. The red points and red fences are the average and stand error of all realizations at each integer time.

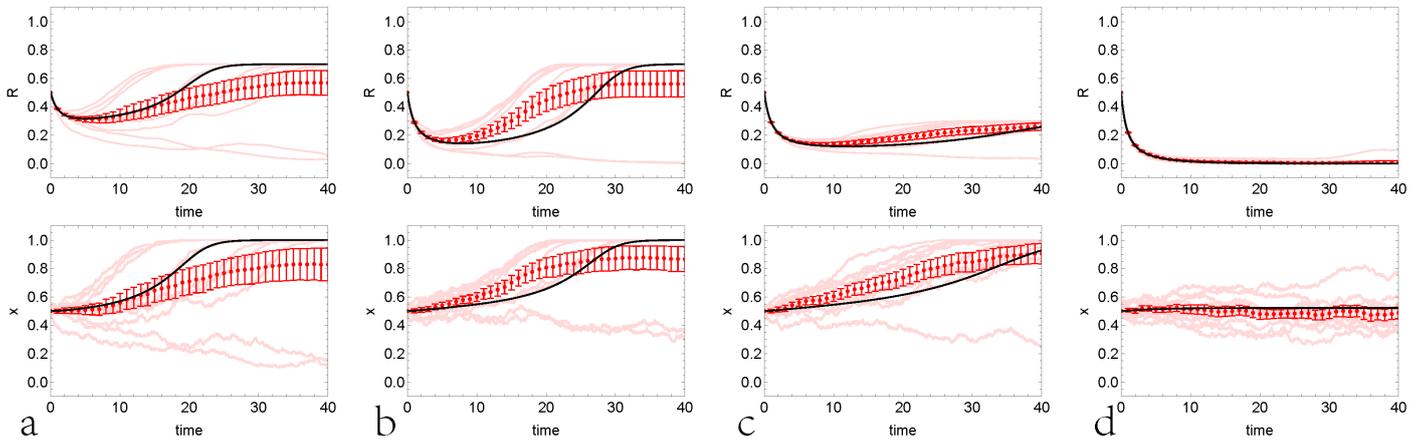

**Supplementary Figure 14 | Comparison between macroscopic ODE and microscopic update for the resource and conformity-driven model with** $c = 0.75$. **(a)** $\hat{e}_C = 0.3, \hat{e}_D = 1.1$; **(b)** $\hat{e}_C = 0.3, \hat{e}_D = 1.5$; **(c)** $\hat{e}_C = 0.7, \hat{e}_D = 1.1$; **(d)** $\hat{e}_C = 0.7, \hat{e}_D = 1.5$. The other parameters are growth rate $T = 2$, player number $N = 500$ and initial condition $R_0 = 0.5, x_0 = 0.5$ as well as independent realization number $n = 10$. Each light red line is one realization of numerical simulation and the black line is the theoretical prediction of macroscopic ODE. The red points and red fences are the average and stand error of all realizations at each integer time.

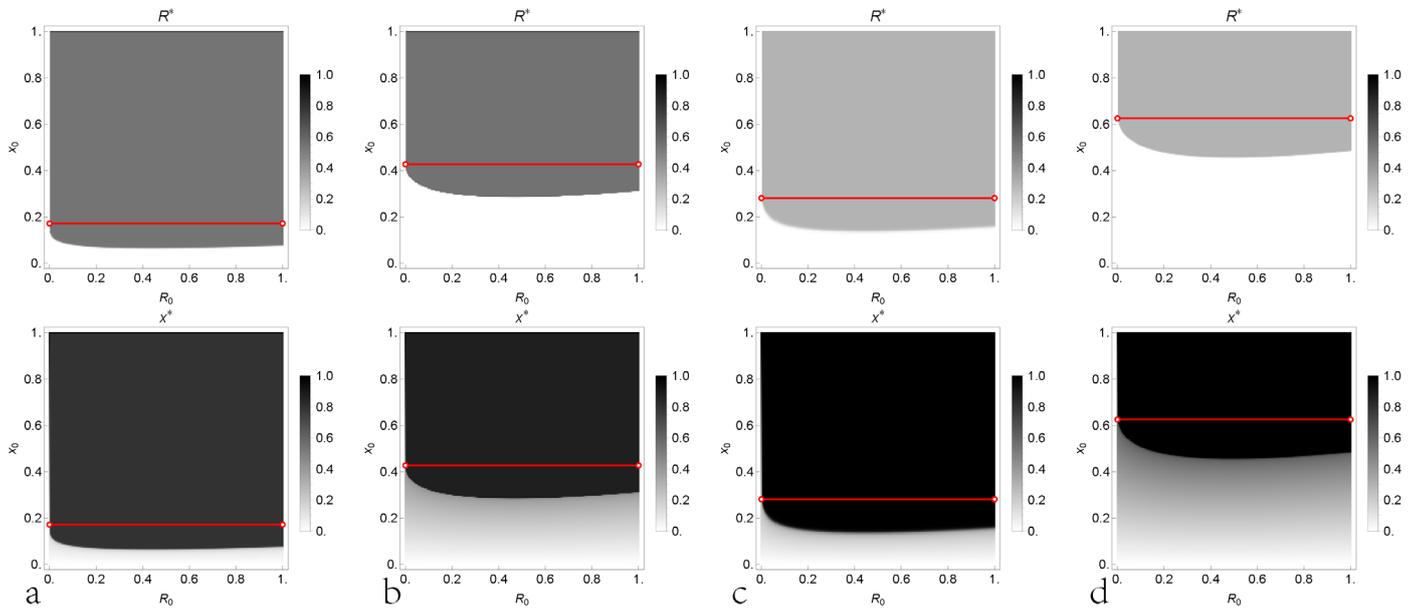

**Supplementary Fig. 15 | Density plot of stable equilibrium** $R^*$ **and** $x^*$ **for the resource and conformity-driven model with** $c = -0.75$. **(a)** $\hat{e}_C = 0.3, \hat{e}_D = 1.1$; **(b)** $\hat{e}_C = 0.3, \hat{e}_D = 1.5$; **(c)** $\hat{e}_C = 0.7, \hat{e}_D = 1.1$; **(d)** $\hat{e}_C = 0.7, \hat{e}_D = 1.5$. The other parameters are growth rate $T = 2$.

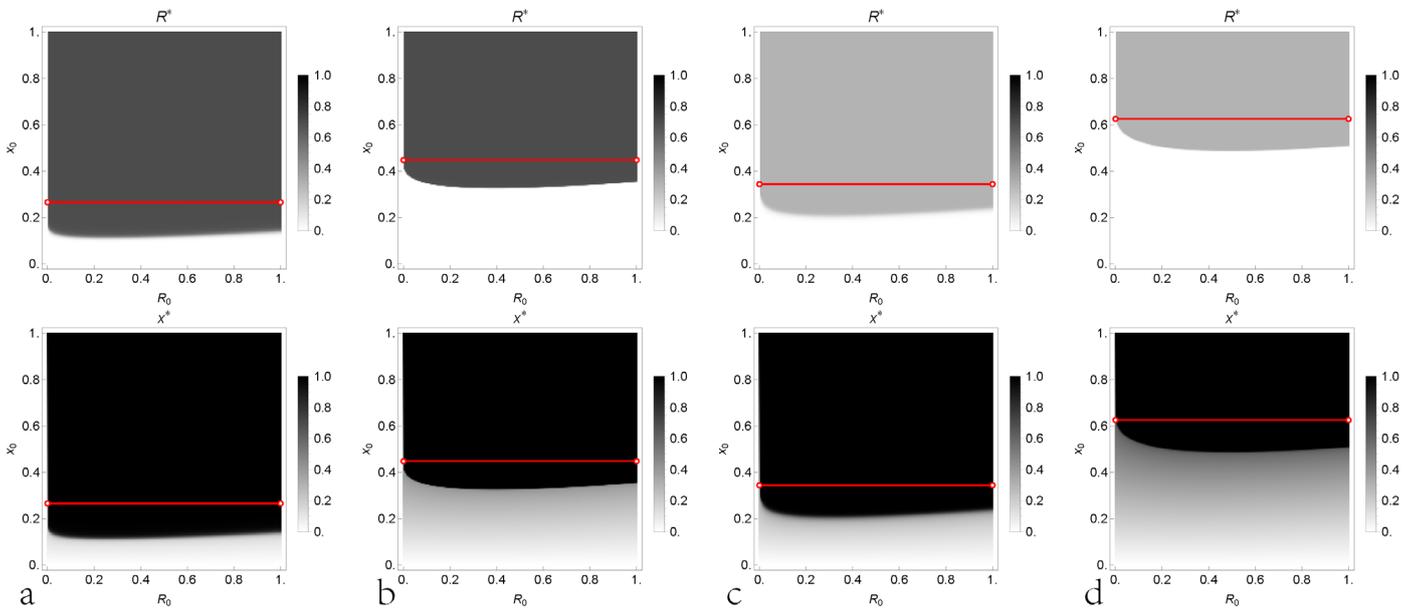

**Supplementary Fig. 16 |** Density plot of stable equilibrium $R^*$ and $x^*$ for the resource and conformity-driven model with $c = -0.25$. **(a)** $\hat{e}_C = 0.3, \hat{e}_D = 1.1$; **(b)** $\hat{e}_C = 0.3, \hat{e}_D = 1.5$; **(c)** $\hat{e}_C = 0.7, \hat{e}_D = 1.1$; **(d)** $\hat{e}_C = 0.7, \hat{e}_D = 1.5$. The other parameters are growth rate $T = 2$.

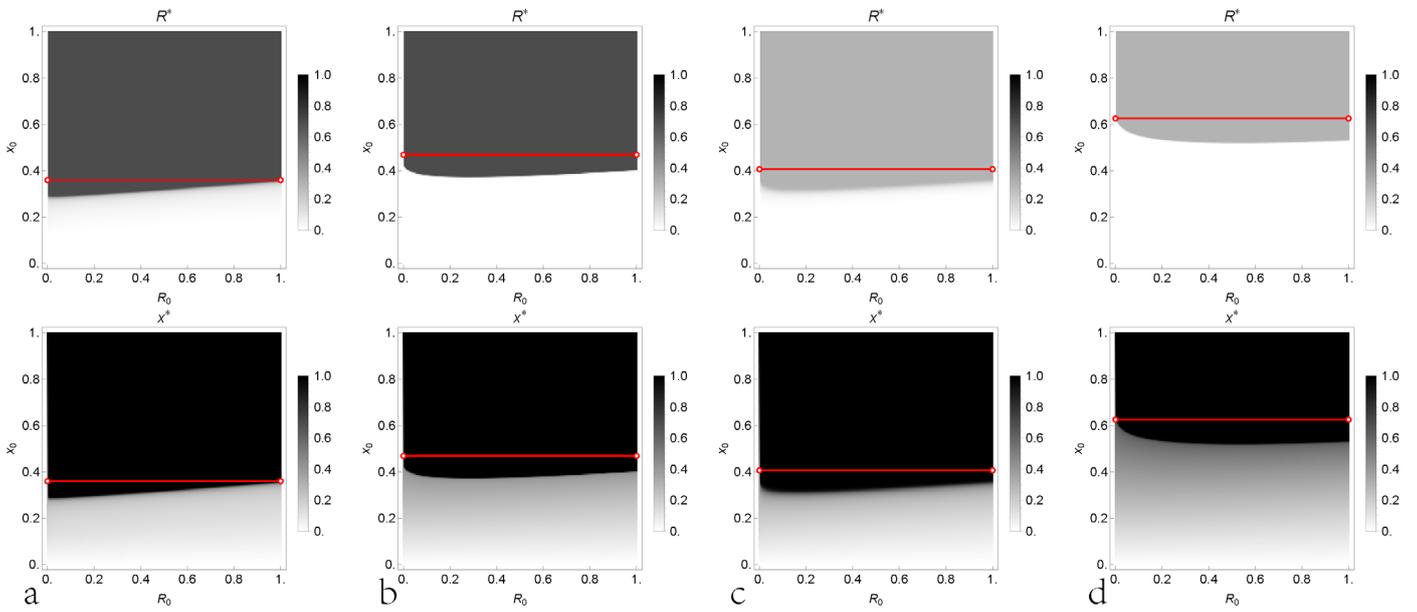

**Supplementary Fig. 17 |** Density plot of stable equilibrium $R^*$ and $x^*$ for the resource and conformity-driven model with $c = 0.25$. **(a)** $\hat{e}_C = 0.3, \hat{e}_D = 1.1$; **(b)** $\hat{e}_C = 0.3, \hat{e}_D = 1.5$; **(c)** $\hat{e}_C = 0.7, \hat{e}_D = 1.1$; **(d)** $\hat{e}_C = 0.7, \hat{e}_D = 1.5$. The other parameters are growth rate $T = 2$.

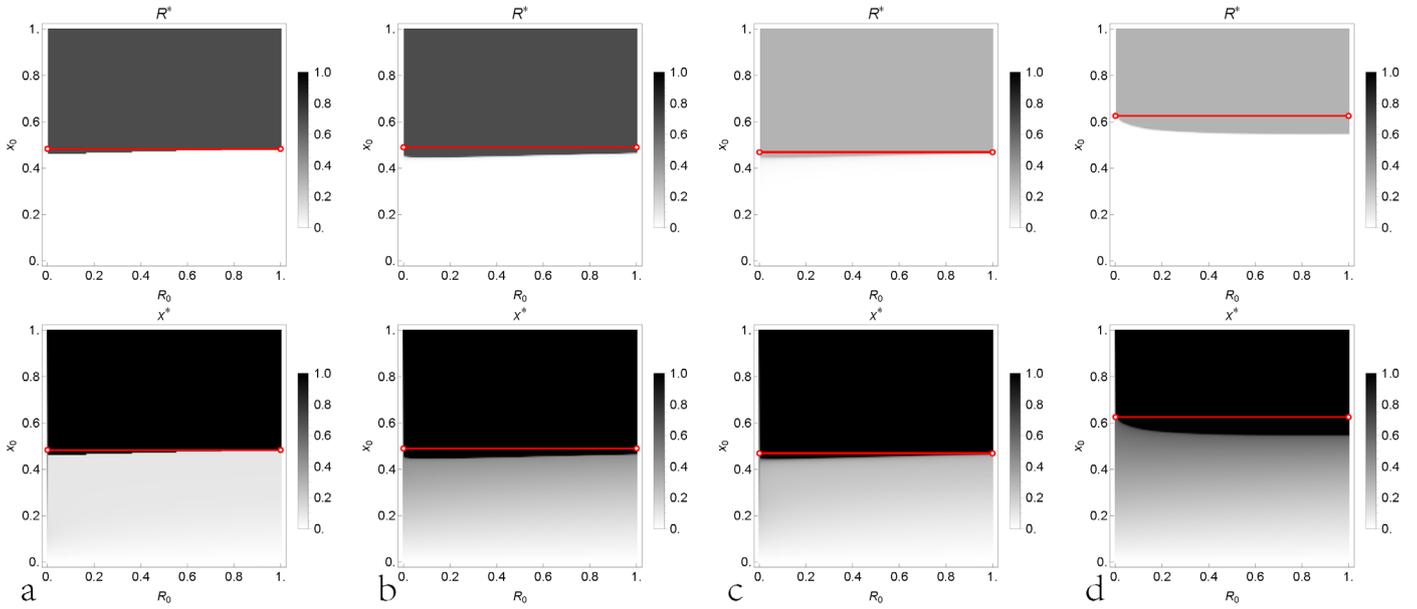

**Supplementary Fig. 18 |** Density plot of stable equilibrium $R^*$ and $x^*$ for the resource and conformity-driven model with $c = 0.75$. **(a)** $\hat{e}_C = 0.3, \hat{e}_D = 1.1$; **(b)** $\hat{e}_C = 0.3, \hat{e}_D = 1.5$; **(c)** $\hat{e}_C = 0.7, \hat{e}_D = 1.1$; **(d)** $\hat{e}_C = 0.7, \hat{e}_D = 1.5$. The other parameters are growth rate $T = 2$.

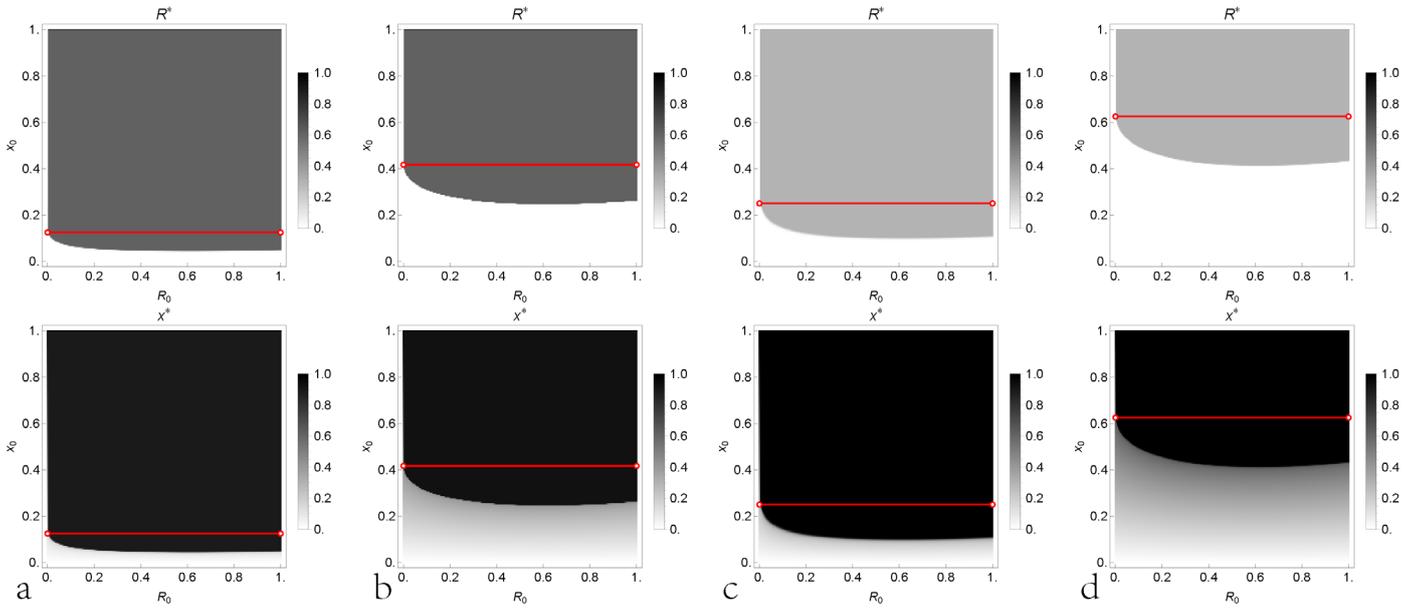

**Supplementary Fig. 19 |** Density plot of stable equilibrium $R^*$ and $x^*$ for the resource-driven model with quadratic form. **(a)** $\hat{e}_C = 0.3, \hat{e}_D = 1.1$; **(b)** $\hat{e}_C = 0.3, \hat{e}_D = 1.5$; **(c)** $\hat{e}_C = 0.7, \hat{e}_D = 1.1$; **(d)** $\hat{e}_C = 0.7, \hat{e}_D = 1.5$. The other parameters are growth rate $T = 2$ as well as greed parameter $w = f(x) = aR^2 + bR + c$ where $a + b = 2, c = -1$.

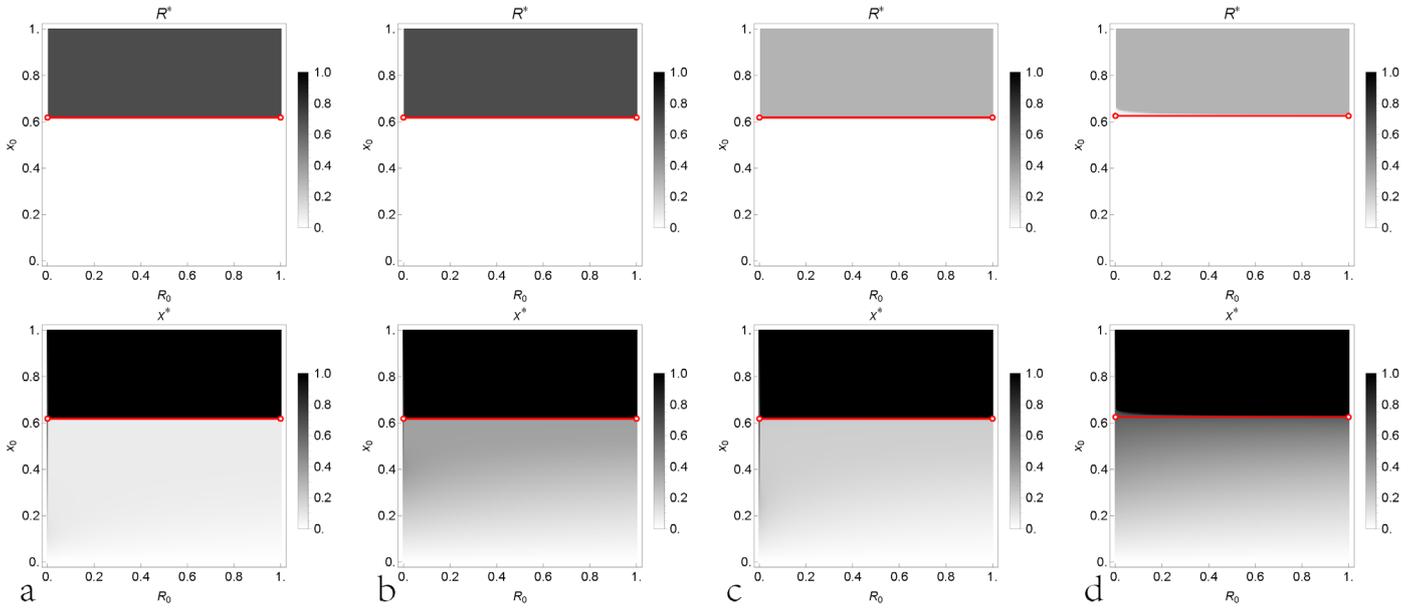

**Supplementary Fig. 20 | Density plot of stable equilibrium** $R^*$ **and** $x^*$ **for the conformity-driven model with quadratic form.** (a) $\hat{e}_C = 0.3, \hat{e}_D = 1.1$; (b) $\hat{e}_C = 0.3, \hat{e}_D = 1.5$; (c) $\hat{e}_C = 0.7, \hat{e}_D = 1.1$; (d) $\hat{e}_C = 0.7, \hat{e}_D = 1.5$. The other parameters are growth rate $T = 2$ as well as greed parameter $w = f(x) = ax^2 + bx + c$ where $a + b = -2, c = 1$.

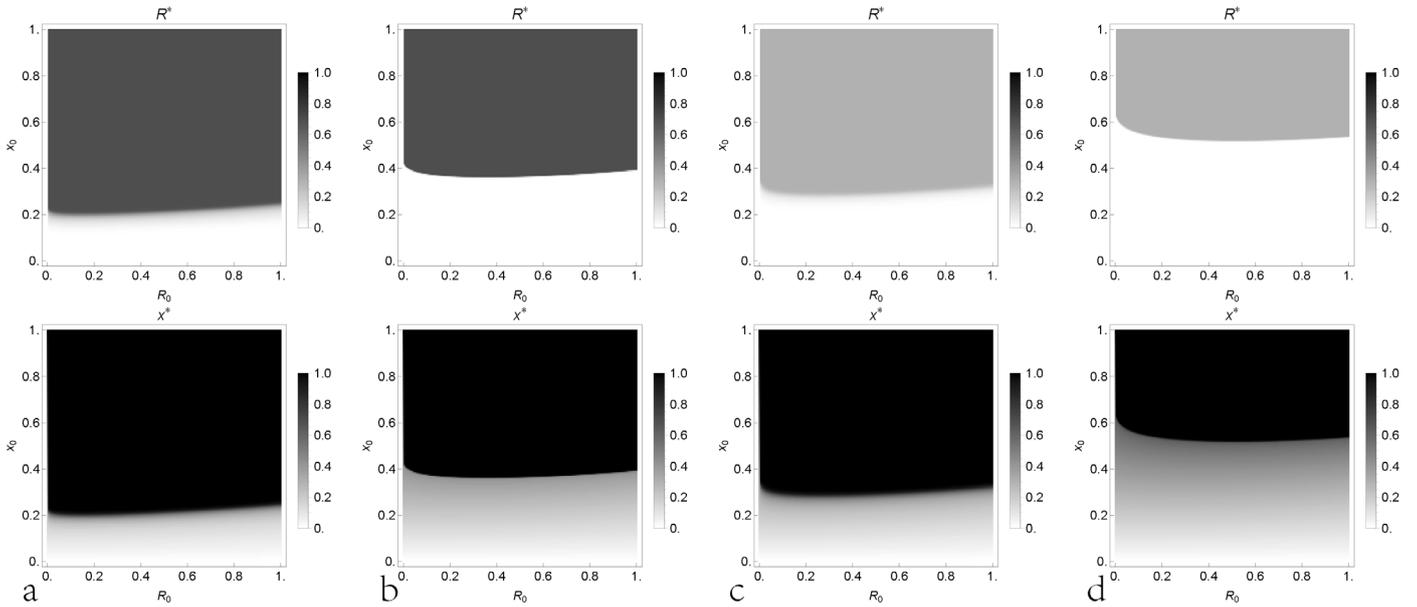

**Supplementary Fig. 21 | Density plot of stable equilibrium** $R^*$ **and** $x^*$ **for the resource and conformity-driven model with quadratic form.** (a) $\hat{e}_C = 0.3, \hat{e}_D = 1.1$; (b) $\hat{e}_C = 0.3, \hat{e}_D = 1.5$; (c) $\hat{e}_C = 0.7, \hat{e}_D = 1.1$; (d) $\hat{e}_C = 0.7, \hat{e}_D = 1.5$. The other parameters are growth rate $T = 2$ as well as greed parameter $w = f(x) = aR^2 + bR + cx^2 + dx + e$ where $a = 0.5, b = 0.5, c = 0.5, d = 0.5, e = 0$.

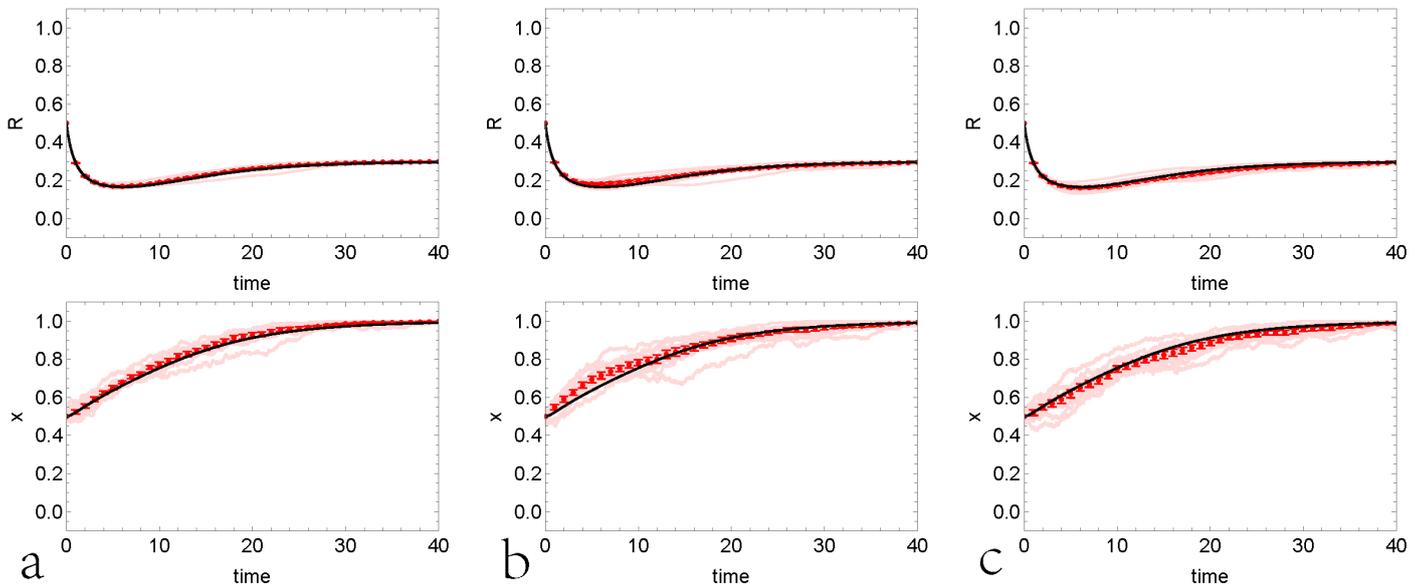

**Supplementary Figure 22 | Comparison between macroscopic ODE and microscopic update for the resource-driven model under varied social networks.** (**a**) complete network; (**b**) Barabasi-Albert network; (**c**) small-world network. The other parameters are growth rate $T=2$, player number $N=500$, normalized extraction rates $\hat{e}_C=0.7, \hat{e}_D=1.1$ and initial condition $R_0=0.5, x_0=0.5$ as well as independent realization number $n=10$. Each light red line is one realization of numerical simulation and the black line is the theoretical prediction of macroscopic ODE. The red points and red fences are the average and stand error of all realizations at each integer time.

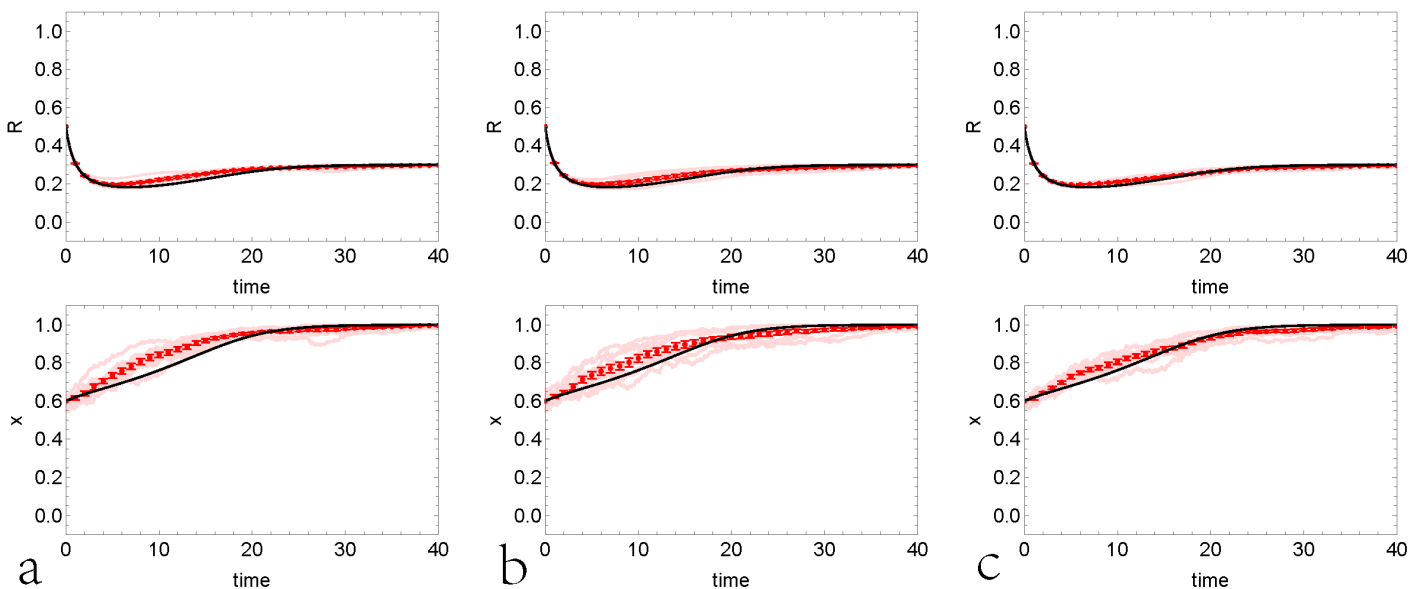

**Supplementary Figure 23 | Comparison between macroscopic ODE and microscopic update for the conformity-driven model under varied social networks.** (**a**) complete network; (**b**) Barabasi-Albert network; (**c**) small-world network. The other parameters are growth rate $T=2$, player number $N=500$, normalized extraction rates $\hat{e}_C=0.7, \hat{e}_D=1.1$ and initial condition $R_0=0.5, x_0=0.5$ as well as independent realization number $n=10$. Each light red line is one realization of

numerical simulation and the black line is the theoretical prediction of macroscopic ODE. The red points and red fences are the average and stand error of all realizations at each integer time.

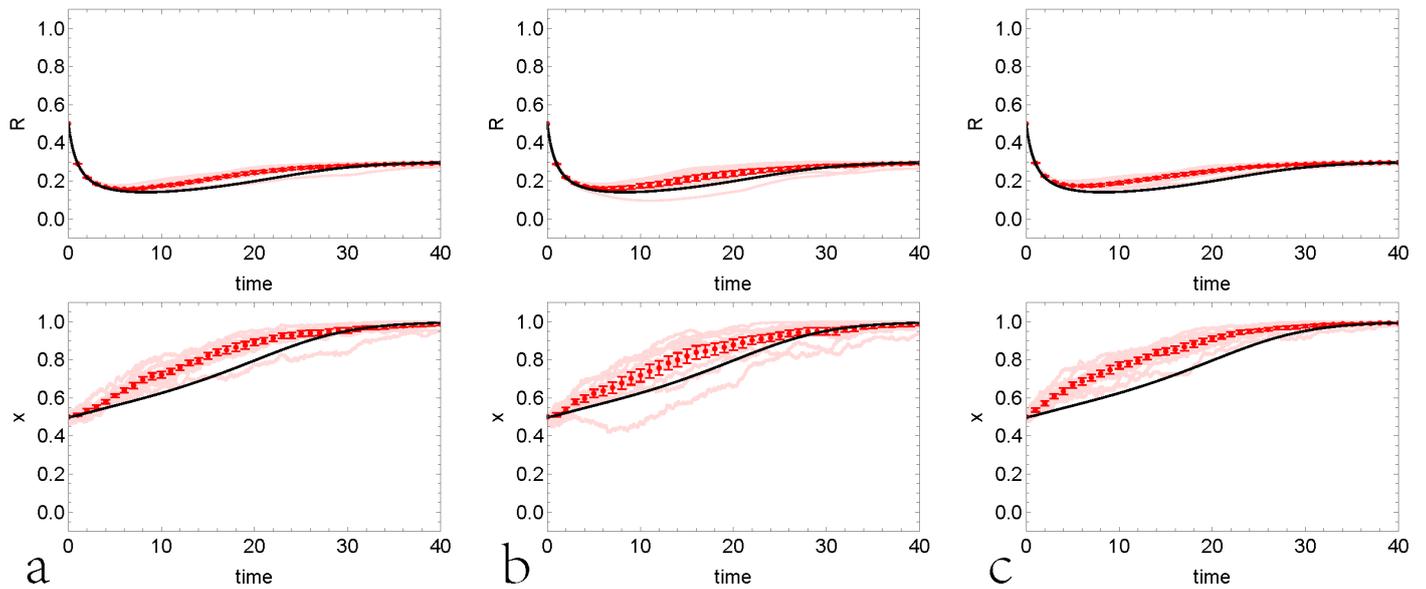

**Supplementary Figure 24 | Comparison between macroscopic ODE and microscopic update for the resource and conformity-driven model with** $c = 0.25$ **under varied social networks.** (**a**) complete network; (**b**) Barabasi-Albert network with connectivity 0.2; (**c**) small-world network with connectivity 0.2. The other parameters are growth rate $T = 2$, player number $N = 500$, normalized extraction rates $\hat{e}_C = 0.7, \hat{e}_D = 1.1$ and initial condition $R_0 = 0.5, x_0 = 0.5$ as well as independent realization number $n = 10$. Each light red line is one realization of numerical simulation and the black line is the theoretical prediction of macroscopic ODE. The red points and red fences are the average and stand error of all realizations at each integer time.